\RequirePackage{fix-cm}
\documentclass[twocolumn, epjc3, comma, sort&compress, natbib]{svjour3}
\bibpunct{[}{]}{,}{n}{}{,} 

\journalname{Eur. Phys. J. A}

\usepackage{latexsym}
\usepackage{amsmath}
\usepackage{amssymb}
\usepackage{amsfonts}
\usepackage{comment}
\usepackage{CJKutf8}

\usepackage[mathscr,scaled=1.15]{urwchancal}
\DeclareFontFamily{OT1}{pzc}{}
\DeclareFontShape{OT1}{pzc}{m}{it}%
{<-> s * [1.15] pzcmi7t}{}
\DeclareMathAlphabet{\mathpzc}{OT1}{pzc}{m}{it}

\usepackage{color}

\usepackage{supertabular}
\usepackage{placeins}
\usepackage{epsfig}
\usepackage{graphicx}

\definecolor{purple}{rgb}{0.5,0,0.5}
\definecolor{blue}{rgb}{0.0,0,0.9}
\definecolor{prdblue}{rgb}{0.133,0.118,0.498}
\usepackage[colorlinks=true, pdfstartview=FitV, linkcolor=prdblue, citecolor= prdblue, urlcolor=prdblue]{hyperref}

\begin{document}
\begin{CJK*}{UTF8}{gbsn}

\title{$\,$\\[-6ex]\hspace*{\fill}{\normalsize{\sf\emph{Preprint nos}.\
NJU-INP 119/26,
USTC-ICTS/PCFT-26-32}}\\[1ex]
Contact interaction treatment of $\pi$ and $\rho$ elastic and transition tensor form factors}

\author{Ke-Yu Guo (郭柯妤)\thanksref{NJU,INP}%
       $\,^{\href{https://orcid.org/0009-0003-1343-5652}{\textcolor[rgb]{0.00,1.00,0.00}{\sf ID}}}$
\and
        Ke-Xin Zeng (曾可欣)\thanksref{NJU,INP}%
        $\,^{\href{https://orcid.org/0009-0001-6575-5645}{\textcolor[rgb]{0.00,1.00,0.00}{\sf ID}}}$
\and
    \\Xin-Yu Bai (白芯瑜)\thanksref{NJU,INP}%
       $\,^{\href{https://orcid.org/0009-0007-4429-103X}{\textcolor[rgb]{0.00,1.00,0.00}{\sf ID}}}$
\and
       Chen Chen (陈晨)\thanksref{ICTS,PHCFT}%
       $\,^{\href{https://orcid.org/0000-0003-3619-0670}{\textcolor[rgb]{0.00,1.00,0.00}{\sf ID}}}$
\and 
    Craig D. Roberts\thanksref{NJU,INP}%
       $\,^{\href{https://orcid.org/0000-0002-2937-1361}{\textcolor[rgb]{0.00,1.00,0.00}{\sf ID}}}$
}

\authorrunning{Ke-Yu Guo \emph{et al}.} 

\institute{School of Physics, Nanjing University, Nanjing, Jiangsu 210093, China \label{NJU}
           \and
           Institute for Nonperturbative Physics, Nanjing University, Nanjing, Jiangsu 210093, China \label{INP}
           \and
           Interdisciplinary Center for Theoretical Study, \href{https://ror.org/04c4dkn09}{University of Science and Technology of China}, Hefei, Anhui 230026, China \label{ICTS}
           \and
            Peng Huanwu Center for Fundamental Theory, Hefei, Anhui 230026, China \label{PHCFT}
\\[1ex]
Email:
\href{mailto:chenchen1031@ustc.edu.cn}{chenchen1031@ustc.edu.cn} (CC);
\href{mailto:cdroberts@nju.edu.cn}{cdroberts@nju.edu.cn} (CDR)
            }

\date{2026 September 30}

\maketitle
\end{CJK*}

\begin{abstract}
Predictions for tensor charges and form factors, elastic and transition, involving $\pi$- and $\rho$-mesons and their scalar and axialvector diquark partners, are delivered using a symmetry-preserving treatment of a vector\,$\otimes$\,vector contact interaction (SCI).
Two distinct SCI regularisation schemes are employed, with the results showing little sensitivity.
Although, as typical in SCI analyses, the form factors are stiff; their infrared behaviour may be considered physically reliable.
Notable amongst related quantities are the following:
the pion tensor charge is approximately $0.36$ and the associated tensor form factor radius is practically the same as the pion charge radius;
the $\rho$-meson tensor charge is roughly 80\% of that for the proton;
and diquark tensor charges and form factors are semiquantitatively alike with those of their $\pi$, $\rho$ partners.
In addition to being interesting in themselves, the SCI predictions can serve as baselines for future studies with a closer connection to QCD.
\end{abstract}


\section{Introduction}
\label{secintro}
The Lagrangian of quantum chromodynamics (QCD) is written in terms of colour-carrying gluon and quark parton fields \cite{Fritzsch:1973pi, Pickering:1984tk}; however, these partonic quanta are not directly observable.
Only colour-singlet bound states seeded by such degrees of freedom (dof) have been ``seen'' in currently achievable detectors.
Seemingly, in terms of theoretical predictions, the simplest bound states are mesons, \emph{i.e}., systems formed from a valence quark parton and a valence antiquark parton.
Amongst mesons, the $\pi$ and $\rho$ each hold a special place:
the $J^{PC}=0^{-+}$ $\pi$ because it is Nature's lightest hadron and most fundamental Nambu-Goldstone (NG) boson \cite{Horn:2016rip, Roberts:2021nhw};
and the $1^{--}$ $\rho$ because, in quark potential models, it is described as being simply related to the $\pi$ by a spin flip excitation of the valence dof \cite[PDG\,-\,Ch.\,15]{ParticleDataGroup:2024cfk}.

QCD provides a deeper understanding of $\pi$, $\rho$ properties.
The NG boson character of the $\pi$ is an emergent feature of strong interactions, owing to dynamical chiral symmetry breaking (DCSB), which is itself a corollary of the dynamics that drives emergent hadron mass (EHM) \cite{Horn:2016rip, Roberts:2021nhw, Ding:2022ows, deTeramond:2022zcm, Ferreira:2023fva, Raya:2024ejx, Achenbach:2025kfx, Binosi:2026tre}.
The pion wave function is complicated: at the hadron scale \cite{Cui:2020tdf, Yin:2023dbw}, viewed from the light-front \cite{Brodsky:1997de, Heinzl:2000ht}, it is found to give roughly equal weight to valence dof helicity antiparallel and parallel components \cite{Xiao:2025cqz}.
Evidently, therefore, the pion is not a simple $S$-wave bound state.

Turning to the $\rho$, in QCD, this state is far more than merely a spin-flip excitation of the $\pi$.
Whilst a complete result for the $\rho$ light-front wave function (LFWF) is currently unavailable, it is known to possess large non-$S$-wave (non-helicity-aligned) components \cite{Gao:2014bca}; and viewed from the rest-frame, the $\rho$-meson's hadron-scale Poin\-car\'e-covariant Bethe-Salpeter wave function involves \mbox{$S$-}, $P$-, and $D$-wave elements \cite{Hilger:2015ora}.
The relative strengths of these components depends on the truncation used for the quark + antiquark scattering kernel, with more realistic approximations yielding larger non-$S$-wave components \cite{Xu:2022kng}.

In this context of in-hadron valence dof orbital angular momentum, hadron elastic and transition tensor form factors are interesting because, from a light-front perspective, their $Q^2=0$ values -- the tensor charges -- measure the relative strength/overlap of different helicity projection components in the LFWFs of the hadrons involved.
This is in contrast to the elastic electric charge form factors -- canonical LFWF normalisations -- whose $Q^2=0$ values are, instead, the sum of the magnitudes-squared of these components.
In general, the tensor form factors can be related to the hadron's leading-twist tensor (transverse-polarisation) generalised parton distribution \cite{Diehl:2003ny, Meissner:2008ay}, but we do not involve that connection herein.

This study focuses on the $\pi$ and $\rho$ elastic tensor form factors and the $\pi\leftrightarrow\rho$ transition  tensor form factor.
In doing so, it complements and extends work described in, \emph{e.g}., Refs.\,\cite{Zhang:2020ecj, Adhikari:2021jrh, Wang:2022mrh, Puhan:2025pfs, Alexandrou:2021ztx}.
As a natural byproduct of the meson study, we also obtain results for kindred tensor form factors that involve the isoscalar-scalar, $0^+$, and isovector-axialvector, $1^+$, quark + quark (diquark) correlations that are  believed to play an important role in baryon structure \cite{Barabanov:2020jvn, Yin:2021uom, Ding:2022ows, Achenbach:2025kfx}.
We formulate the computational problem using continuum Schwinger function methods (CSMs) \cite{Eichmann:2016yit, Fischer:2018sdj, Huber:2018ned, Qin:2020rad} and work in the ${\cal G}$-parity symmetry limit throughout; hence, \emph{e.g}., in a positively charged meson, the valence $u$ and $\bar d$ form factors are identical.

Tensor charges are not conserved; so, their value depends on the resolving scale, $\zeta$.
Our analysis delivers initial results for tensor form factors at the hadron scale \cite{Cui:2020tdf, Yin:2023dbw}: $\zeta_{\cal H}=0.33\,$GeV.
All-orders (AO) evolution \cite{Cui:2020tdf, Yin:2023dbw} is then used to define them at any $\zeta>\zeta_{\cal H}$.
Notably, however, normalised by the associated tensor charge $(Q^2=0$ value), a given tensor form factor is scale invariant.
This is a direct consequence of multiplicative renormalisability, as may be seen, \emph{e.g.}, by adapting the discussion in Ref.\,\cite[Sect.\,5.1]{Cui:2020tdf} to a comparative analysis of pion elastic charge and tensor form factors.

In CSM analyses of meson properties, the basic element is the approximation used for the quark + antiquark scattering kernel.
This choice involves two steps.
The first identifies which trajectory through interaction space will be summed.
A systematic, symmetry-preserving scheme for identifying and following this path was introduced in Refs.\,\cite{Munczek:1994zz, Bender:1996bb}.
The leading-order trajectory is called rainbow-ladder (RL) truncation, which is employed herein. 
In the second step, one refers to studies of QCD's gauge sector and extracts the process-independent interaction that will, for instance, support bound-state formation.

A simple and widely used approach is the symmetry-pre\-ser\-ving treatment of a vector\,$\otimes$\,vector contact interaction (SCI).
Introduced in Refs.\,\cite{Gutierrez-Guerrero:2010waf, Roberts:2011wy}, the SCI provides an insightful tool for the study of hadron observables, which enables baselines to be drawn for more sophisticated studies.
For instance, comparisons between SCI results and predictions obtained using QCD-con\-nec\-ted interactions \cite{Qin:2011dd, Binosi:2014aea} serve to reveal a given observable's sensitivity to the pointwise behaviour of the quark-quark interaction and phenomena associated with the emergence of hadron mass
\cite{Horn:2016rip, Roberts:2021nhw, Ding:2022ows, deTeramond:2022zcm, Ferreira:2023fva, Raya:2024ejx, Achenbach:2025kfx, Binosi:2026tre}.

In the sixteen years since its first use, the SCI has been refined.
It cannot be a precision tool; notwithstanding, today's formulation has many strengths, such as:
algebraic simplicity;
simultaneous applicability to a large array of systems and processes;
capacity for supplying insights that connect and explain a wide variety of phenomena;
and utility as a means of evaluating the viability of algorithms used in calculations that otherwise depend upon high performance computing.
Contemporary SCI applications are normally parameter-free and numerous benchmarking predictions have already been obtained for phenomena involving mesons; see, \emph{e.g}., Refs.\,\cite{Roberts:2011wy, Chen:2012txa, Serna:2017nlr, Zhang:2020ecj, Xing:2022sor, Sultan:2024hep, Xing:2025eip, Gutierrez-Guerrero:2019uwa, Chen:2024emt, Cheng:2026nud, Gutierrez-Guerrero:2026rsb}.
Baryon studies are also available \cite{Gutierrez-Guerrero:2019uwa, Chen:2024emt, Segovia:2013rca, Xu:2015kta, Raya:2021pyr, Yin:2021uom, Cheng:2022jxe, Yu:2025fer, Bai:2026nqo}.

This report is organised as follows.
The SCI form of the dressed quark-tensor vertex is calculated and discussed in Sect.\,\ref{QTV}, which also explains tensor charge scale evolution and provides an estimate of $u$ and $d$ valence dof contributions to the nucleon tensor charge.  (\ref{AppendixSCI} provides all information necessary to understand and repeat each SCI calculation herein.)
Section~\ref{sec2} describes SCI predictions for the pion tensor form factor.
Therein and thereafter, we present results obtained with two distinct regularisation schemes.
In doing so, we provide one indication of the systematic error in our SCI analyses.  (It is usually small.)
A comparison is also made with some results from the numerical simulation of lattice-regularised QCD (lQCD).
The five elastic tensor form factors of the $\rho$-meson are discussed in Sect.\,\ref{sec4}, with reliable interpolations thereof recorded in \ref{InterprhoT}.
Section~\ref{sec5} reports SCI predictions for the three $\pi \leftrightarrow \rho$ tensor transition form factors, 
interpolations of which are provided in \ref{ApppirhoI}.
The mapping of $\pi$, $\rho$ tensor form factors into kindred $0^+$, $1^+$ diquark form factors is explained in Sect.\,\ref{sec6}, with interpolations recorded in Appendix~\ref{AppE}.
Section~\ref{epilogue} presents a summary and perspective.

\section{Quark tensor vertex}
\label{QTV}
A prerequisite for any CSM analysis of a hadron's tensor form factor(s) is calculation of the dressed quark-tensor vertex.  In our RL SCI formulation of the problem, this vertex, $\Gamma_{\mu \nu}^{\mathrm{T}}$, satisfies the following inhomogeneous Bethe-Salpeter equation ($l_+=l+Q$):
\begin{align}
\Gamma_{\mu \nu}^{\mathrm{T}}&(Q )  =\sigma_{\mu \nu} \nonumber \\
& \quad +
C_2(R) \frac{4 \pi \alpha_\mathrm{IR}}{m_{G}^{2}} \int_{dl} \gamma_{\alpha}  S(l_+) i\Gamma_{\mu \nu}^{\mathrm{T}}(Q) S(l) i\gamma_{\alpha},
\label{TVbse}
\end{align}
where
$C_2(R) = (N_c^2-1)/(2N_c)$, $N_c = 3$, and the other multiplicative factors are specified in \ref{SCIback};
$\int_{dl}$ indicates a symmetry-preserving regularisation of the four dimensional integral;
and $S$ is the dressed light-quark propagator, \ref{GapEq}.

Since tensor charges are not conserved, $\Gamma_{\mu \nu}^{\mathrm{T}}$ depends on the resolving scale, $\zeta$.
(For notational simplicity, we do not make this dependence explicit.)
Indeed, in principle, every element in the Eq.\,\eqref{TVbse} integrand depends on $\zeta$; however, such dependence is a matter of definition when using the SCI because this framework does not represent a renormalisable theory: for phenomenological purposes/comparisons, as has long been common \cite{Jaffe:1980ti}, QCD evolution is imposed \cite{Dokshitzer:1977sg, Gribov:1971zn, Lipatov:1974qm, Altarelli:1977zs}.
Notably, rather than introducing an enumerable infinity of counterterms, as typical in effective field theories, an ultraviolet cutoff is used and made meaningful in formulating the SCI; see Eq.\,\eqref{SCIReg1}.
Thereafter, SCI results are defined to be valid at the hadron scale: $\zeta_{\cal H}=0.33\ \mathrm{GeV}\approx M$, where $M$ is the light-quark dressed mass, a principal signature of EHM; see \ref{GapEq}.
(The value of $\zeta_{\cal H}$ is a CSM prediction; see, \emph{e.g}., Refs.\,\cite[Fig.\,1]{Cui:2020tdf}, \cite{Yin:2023dbw}.)

The SCI solution of Eq.\,\eqref{TVbse} has the following form:
{\allowdisplaybreaks
\begin{align}
    \Gamma_{\mu \nu}^{\rm T}(Q ) & =\mathcal{V}_{1}(Q^{2}) \sigma_{\mu \nu}
    +\mathcal{V}_{2}(Q^{2}) \frac{1}{M}[i\gamma\cdot Q,\sigma_{\mu \nu}]\nonumber \\
    & \quad +\mathcal{V}_{3}(Q^{2})\frac{1}{M^2} i\gamma\cdot Q \sigma_{\mu \nu} i\gamma\cdot Q .
\label{TVbsef}
\end{align}
(Its total angular momentum decomposition is discussed in \ref{TensorV}.)
Inserting Eq.\,\eqref{TVbsef} into Eq.\,\eqref{TVbse}, one finds
\begin{subequations}
\label{TensorAnswer}
\begin{align}
    \mathcal{V}_{1}(Q^{2})&=1, \label{V1}\\
    \mathcal{V}_{2}(Q^{2} )&=\frac{{K}_{\cal V}(Q^{2})}{1+K_{\gamma}(Q^{2})}, \label{V2}\\
    \mathcal{V}_{3}(Q^{2} )&=0, \label{V3}
\end{align}
\end{subequations}
where $(\check \alpha = 1-\alpha)$
\begin{subequations}
\begin{align}
{K}_{\cal V}(Q^{2}) & =-\frac{\alpha_\mathrm{IR} M^{2}}{3 \pi m_{G}^{2}} \int_{0}^{1} d \alpha \bar{\mathcal{C}}_{1}^{\rm iu}(w(\alpha,Q^{2}))\,,
\label{KV}\\
K_{\gamma}(Q^{2}) & =\frac{4 \alpha_{\mathrm{IR}}}{3 \pi m_{G}^{2}} \int_{0}^{1} d \alpha \alpha \check\alpha
Q^{2} \bar{\mathcal{C}}_{1}^{\mathrm{iu}}(\omega(\alpha, Q^{2}))\,,\\
    \label{kgamma}
\omega(\alpha, Q^{2}) & =M^{2} + \alpha \check\alpha Q^{2}\,.
\end{align}
\end{subequations}
}

Notably, $\mathcal{V}_{1}(Q^{2})\equiv 1$, independent of $Q^2$.
It follows that the SCI dressed light-quark tensor charge is unity at $\zeta_{\cal H}$:
\begin{equation}
\delta_T^{\zeta_{\cal H}} u = 1 = \delta_T^{\zeta_{\cal H}}d \,.
\label{zHdeltaq}
\end{equation}

It is worth mentioning that in studies with an interaction that better captures the momentum-dependence of QCD's gauge sector,
$\mathcal{V}_{1}(Q^{2}=0) \approx 2/3$ and this function grows toward unity with increasing $Q^2$, being practically ``$1$'' on $Q^2 \gtrsim 2 m_p^2$, where $m_p$ is the proton mass; see, \emph{e.g}., Refs.\,\cite{Yamanaka:2013zoa, Wang:2018kto}.
Thus Eq.\,\eqref{V1} is a fair approximation.
Moreover, regarding $\mathcal{V}_{3}(Q^{2} )$, such analyses predict, like Eq.\,\eqref{V3}, that this form factor is negligible.

Turning to $\mathcal{V}_{2}(Q^{2})$, this function falls uniformly in magnitude as $1/Q^2$ from $\mathcal{V}_{2}(0)=-0.18$.
Hence, whilst $\mathcal{V}_{2}(Q^{2})$ provides a modest contribution to a hadron's tensor charge, its impact on form factors diminishes rapidly with increasing spacelike momentum transfer.
%
Furthermore, combining Eq.\,\eqref{V2} with Eq.\,\eqref{rhoren}, one sees that $\mathcal{V}_{2}(Q^{2} )$ exhibits a pole at the $\rho$-meson mass.  
(Its origin is explained following Eq.\,\eqref{TTLL}.)
This has an effect on most tensor radii.
These $\mathcal V_2$-out\-comes are also consistent with those obtained using realistic mo\-men\-tum-dependent interactions \cite{Yamanaka:2013zoa, Wang:2018kto}.

Since the SCI does not support relative momentum between meson valence dof, there is no pole in the quark-tensor vertex associated with a tensor ($J^P=2^+$) meson \cite{Roberts:2011wy}; see Eqs.\,\eqref{V1}, \eqref{V3}.

All above results and remarks are the same in both SCI regularisation schemes considered herein \cite{Gutierrez-Guerrero:2010waf, Xing:2022jtt}.

As already indicated, we interpret SCI results as being valid at the hadron scale.
Following Ref.\,\cite{Pitschmann:2014jxa}, evolution is imposed and implemented via the AO scheme \cite{Yin:2023dbw}, using which the ratio of quark current masses at any two scales is given by
\begin{equation}
\frac{m^{\zeta_2}}{m^{\zeta_1}}
= \bigg[ \exp\int_{\zeta_1^2}^{\zeta_2^2} \frac{ds}{s}\, \hat\alpha(s)/(\gamma_m \pi) \bigg]^{\gamma_m}\,,
\label{AOmass}
\end{equation}
where $\gamma_m=12/(11 N_c - 2 N_f)$ is the quark mass anomalous dimension
and $\hat\alpha(s)$ is the QCD process-indepen\-dent effective charge, defined, calculated, and discussed in Refs.\,\cite{Cui:2019dwv, Deur:2023dzc, Brodsky:2024zev}.
(See Ref.\,\cite[Eq.\,(13)]{Cui:2019dwv} for a reliable interpolation; the number of active flavours is $N_f=4$.)
Using Eq.\,\eqref{AOmass}, the hadron scale current masses in Table~\ref{Tab:DressedQuarks} correspond to ($\zeta_2=2\,$GeV):
\begin{equation}
m_{u,d}^{\zeta_2} = 0.003\,{\rm GeV}\,,
\quad
m_s^{\zeta_2} = 0.070\,{\rm GeV}\,;
\end{equation}
values which are a fair match with those typically quoted \cite[PDG]{ParticleDataGroup:2024cfk}.

The analogous expression for the tensor charge of a quark, $q$, is
\begin{equation}
\frac{\delta_T^{\zeta_2}q}{\delta_T^{\zeta_1}q}
= \bigg[\exp \int_{\zeta_1^2}^{\zeta_2^2} \frac{ds}{s}\, \hat\alpha(s)/(\gamma_m \pi) \bigg]^{\eta_T} ,
\end{equation}
where $\eta_T = \gamma_m/3$ is the appropriate anomalous dimension.  Hence, using Eq.\,\eqref{zHdeltaq}, one finds $\delta_T^{\zeta_{2}} u = 0.76$.
It is fair to ask for an uncertainty in this prediction.  To that end, we follow Ref.\,\cite{Pitschmann:2014jxa}. Namely, recognising that modern SCI treatments typically yield results for low momentum transfer observables which are practically indistinguishable from those obtained using more sophisticated interactions \cite{Qin:2011dd}, and such studies have an overall uncertainty that is less-than 15\%, then we attach a conservative 15\% uncertainty to this result:
\begin{equation}
\delta_T^{\zeta_{2}} u = 0.76(11)\,.
\label{deltaTz2}
\end{equation}

Working with the value in Eq.\,\eqref{deltaTz2} and supposing that the proton is described by an SU$(4)$-like quark model wave function \cite{He:1994gz}, then one would find the in-proton tensor charges listed in column~1 below:
\begin{equation}
\begin{array}{c|ccc}
 & {\rm herein} &3\textrm{-body} & \mathrm{uwa} \\\hline
\delta_T^{\zeta_2} u & \phantom{-}\tfrac{4}{3}\delta_T^{\zeta_{2}}= \phantom{-}1.01(15)
& \phantom{-} 0.91(04) & \phantom{-}0.79(02) \\
\delta_T^{\zeta_2} d & -\tfrac{1}{3}\delta_T^{\zeta_{2}}= - 0.25(04) & -0.22(01) & -0.22(01)\\
\end{array}\,.
\end{equation}
The remaining columns are
the CSM three-body Faddeev equation result from Ref.\,\cite{Wang:2018kto}
and an uncertainty weighted average of continuum and lattice results from
Refs.\,\cite{Wang:2018kto, Xu:2015kta, Bhattacharya:2015wna, Abdel-Rehim:2015owa, Alexandrou:2024awx}.
Considering these results, we conclude that extant phenomenological inferences  \cite{Ye:2016prn} underestimate $\delta_T^{\zeta_2} u$ by as much as a factor of two.

\section{Pion Tensor Form Factor}
\label{sec2}
Considering either valence dof in a positively charged pion, using RL truncation and the SCI, the associated (dimensionless) tensor form factor, $F_T^{\pi}$, can be determined as follows:
\begin{align}
\Lambda_{\mu\nu}^{\pi u} (K,Q;\zeta_{\cal H})
& =: T_{\mu\nu}^\pi(K,Q) F_T^{\pi u}(Q^2;\zeta_{\cal H})\,,
\end{align}
where, with $P_{\sf i}$, $P_{\sf o}$ being the incoming and outgoing momenta of the target pion,
$K=(P_{\sf o}+P_{\sf i})/2$, $Q=P_{\sf o} - P_{\sf i}$, $P_{\sf i}^2=-m_\pi^2=P_{\sf o}^2$,
\begin{equation}
T_{\mu\nu}^\pi(K,Q)=[Q_\mu K_\nu - K_\mu Q_\nu]/m_\pi\,,
\label{Tpinorm}
\end{equation}
and
\begin{align}
\Lambda_{\mu\nu}^{\pi u} & (K,Q;\zeta_{\cal H}) =
2 N_{c} \operatorname{tr}_{\mathrm{D}} \int_{dq}[i\bar{\Gamma}^{\pi}(-P_{\sf o}) \nonumber \\
& \times S(q+P_{\sf o}) i \Gamma_{\mu \nu}^{T}(Q )  S(q+P_{\sf i}) i \Gamma^{\pi}(P_{\sf i}) S(q)].
\label{IApion}
\end{align}
In Eq.\,\eqref{IApion}:
the trace is over spinor indices and
$\Gamma^{\pi}$ is the canonically normalised pion Bethe-Salpeter amplitude, \ref{MesonBSA}.
As noted above, in QCD, every element in the integrand depends on the resolving scale, $\zeta$; hence, so do the tensor form factors.
On the other hand, rescaled by their values at $Q^2=0$, all resulting tensor form factors are scale independent.
It is worth reiterating that we work in the ${\cal G}$-parity symmetry limit, so the $\bar d$-in-$\pi^+$ and $u$-in-$\pi^+$ results are identical, with analogous statements for the $\pi^-$; hence, hereafter, we drop the superscript $u$.

\begin{figure}[t]
\centerline{%
\includegraphics[clip, width=0.45\textwidth]{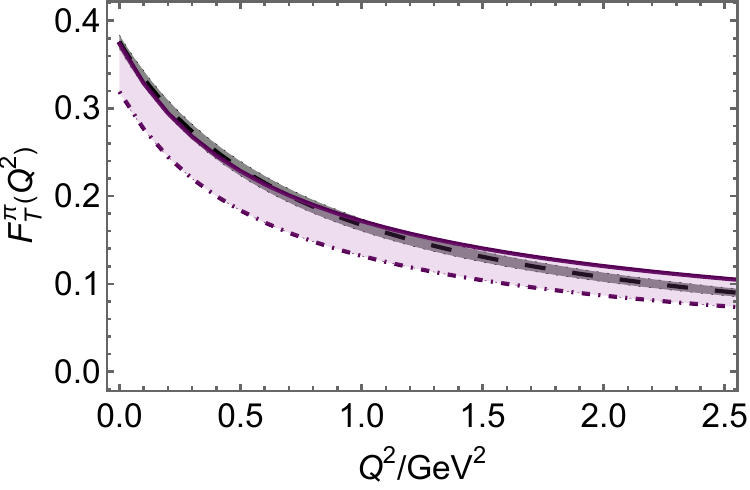}}
\caption{\label{FigpionT}
SCI results for pion tensor form factor at $\zeta=\zeta_2$.
Legend.
Solid purple -- regularisation Scheme I;
dot-dashed purple -- regularisation Scheme II;
long-dashed black comparison curve -- drawn from the lattice-regularised QCD (lQCD) simulation in Ref.\,\cite{Alexandrou:2021ztx}, which used $m_\pi^{\rm lQCD} = 0.265\,$GeV.
}
\end{figure}

\begin{table}[t]
\caption{ \label{TensorCharge}
{\sf Panel A}.
Calculated values for pion $u$ valence-dof tensor charge at resolving scale $\zeta=\zeta_2$ and the $\zeta$-independent tensor charge radius.
For reference, the empirical pion electric charge radius is $0.640(7)\,$fm \cite{Cui:2022fyr};
and existing SCI analyses give
$r_{\rm ch}^{\pi\,{\rm SCI}} \approx 0.56\,$fm; see Eq.\,\eqref{Fpiem}.
%
{\sf Panel B}. Analogues for the isoscalar-scalar diquark.  Here, no lQCD result is available.
%
%
}
\begin{tabular*}
{\hsize}
{
l@{\extracolsep{0ptplus1fil}}
|l@{\extracolsep{0ptplus1fil}}
l@{\extracolsep{0ptplus1fil}}
l@{\extracolsep{0ptplus1fil}}
}\hline\hline
\centering
 {\sf A}                    & SCI-I\; & SCI-II \; & \mbox{\;\cite[lQCD]{Alexandrou:2021ztx}\;}
\\\hline
$m_\pi/{\rm GeV}\ $ & $0.14\ $ & $0.14\ $ & $0.265\ $ \\
$\kappa_T^\pi\ $ & $0.37\ $&  $0.32\ $ & $0.38(01)\ $ \\
$r_T^\pi/{\rm fm}\ $ & $0.57\ $&  $0.60\ $ & $0.67(10)\ $  \\\hline\hline
{\sf B}                    & SCI-I & SCI-II & \\\hline
$m_{0^+}/{\rm GeV}\ $ & $0.78\ $ & $0.78 $ &  \\
$\kappa_T^{0^+}\ $ & $1.72\ $&  $1.76\ $ & \\
$r_T^{0^+}/{\rm fm}\ $ & $0.60\ $&  $0.62\ $ &  \\\hline\hline
\end{tabular*}
\end{table}
One should recognise that the $1/m_\pi$ factor in $T_{\mu\nu}^\pi$ is merely conventional.
We use it to make contact with other studies.  In our view, it would be better to normalise using the pion leptonic decay constant, $f_\pi$, which, as an order parameter for DCSB, remains nonzero in the chiral limit.
With physical quark current masses, $m_\pi \approx 1.5\,f_\pi$.

Explicit forms for the pion tensor form factor, obtained using both regularisation schemes, are given in \ref{ApppionFT}.  Numerical results are depicted in Fig.\,\ref{FigpionT}.
In displaying results obtained with both regularisation schemes, we provide an indication of the systematic error in this SCI analysis.
These differences vanish in $F_T^\pi$ if the pseudovector part of the pion Bethe-Salpeter amplitude is ignored, \emph{i.e}., if one artificially sets $F_\pi = 0$ in Eq.\,\eqref{PSBSAA}.
(Such would correspond to a typical Nambu--Jona-Lasinio model treatment.)

Figure~\ref{FigpionT} also displays a result for $F_T(Q^2,\zeta_2)$ obtained via the numerical simulation of lattice-regularised QCD (lQCD) \cite{Alexandrou:2021ztx}.
Notably, the result is practically indistinguishable from the SCI prediction.

The lQCD simulation used a pion mass-squared that is roughly 3.6-times larger than experiment.
Owing to the conventional normalisation prescription, Eq.\,\eqref{Tpinorm}, the magnitude of the tensor form factor has a nontrivial dependence on the pion mass.
However, ascertaining the precise behaviour is difficult because an inflated pion mass also means an inflated $\rho$-meson mass.
The location of the $\rho$-meson pole in the quark-tensor vertex, Eq.\,\eqref{TVbsef}, has a material influence on $F_T^\pi$ via $\mathcal{V}_{2}(Q^{2})$: decreasing $m_\pi$ acts to suppress $F_T^\pi(0)$, but it also brings the $\rho$ mass closer to $Q^2=0$ and this works to enhance $F_T^\pi(0)$; see Eq.\,\eqref{AlgebraFpiT}.
We therefore make no attempt to estimate the impact on Ref.\,\cite{Alexandrou:2021ztx} of taking $m_\pi$ to the physical value.

The pion $u$ valence-dof tensor charge is defined as $\kappa_T^\pi = F_T^\pi(Q^2=0)$.  In Table~\ref{TensorCharge}\,A, we list SCI predictions for this charge and compare them with a lQCD result \cite{Alexandrou:2021ztx}.
Evidently, theory favours a value 
\begin{equation}
\kappa_T^\pi(\zeta_2) \approx 0.36(3)\,.
\end{equation}
A consistent result was obtained in Ref.\,\cite{Wang:2022mrh}.

\begin{figure}[t]
\centerline{%
\includegraphics[clip, width=0.45\textwidth]{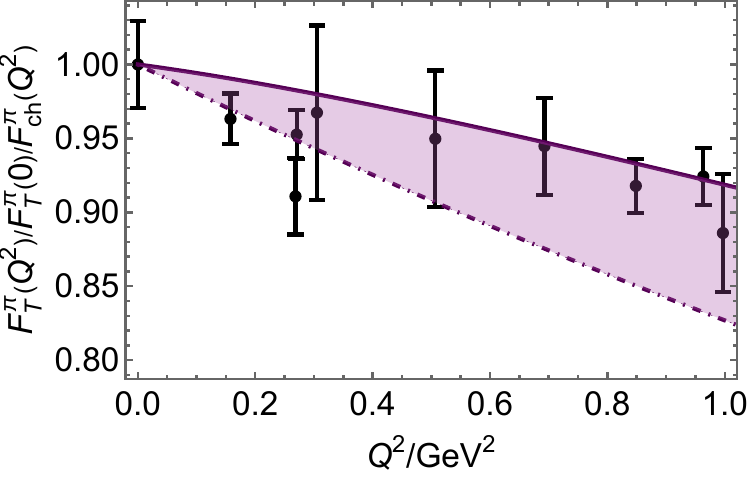}}
\caption{\label{FTonV}
SCI results for the ratio $F_T^\pi(Q^2)/F_T^\pi(0) /F_{\rm ch}^\pi(Q^2)$.
Legend.
Solid purple -- Scheme I;
dot-dashed purple -- Scheme II;
black points -- lQCD results from Ref.\,\cite[Fig.\,8]{Alexandrou:2021ztx}, rescaled so that the central value of the $Q^2=0$ point is unity.
}
\end{figure}

A resolving-scale-independent tensor charge radius can also be defined:
\begin{equation}
(r_T^\pi)^2 = -6 \frac{d}{dQ^2} \ln  F_T^\pi(Q^2) \bigg|_{Q^2=0}\,.
\end{equation}
Again in Table~\ref{TensorCharge}\,A, we list SCI predictions for this radius and compare them with an available lQCD result \cite{Alexandrou:2021ztx}.
The results suggest that the pion tensor and electric radii are very similar.

This outcome is consistent with the analysis in Ref. \cite{Hoferichter:2018zwu};
namely, since both form factors connect to the $\pi \pi$ rescattering channel, then elastic unitarity arguments suggest that $F_T^\pi(Q^2)/F_T^\pi(0) \approx F_{\rm ch}^\pi(Q^2)$ on a domain which extends to $Q^2\lesssim 1\,$GeV$^2$.
We depict SCI predictions for this ratio in Fig.\,\ref{FTonV}, using
Eq.\,\eqref{Fpiem} for the pion charge form factor in both cases, and compare them with lQCD results from Ref.\,\cite[Fig.\,8]{Alexandrou:2021ztx}.
Both SCI and lQCD results agree in indicating that the approximation is fair, but not quantitatively precise.

To close this section, we report interpolations of the SCI results that are reliable on the domain $0\leq z=Q^2\leq 10\,$GeV$^2$:
{\allowdisplaybreaks
\begin{subequations}
\begin{align}
F_T^{\pi {\rm I}}(z) & =0.374 \frac{1 + 0.401 z+0.000725 z^2}{1+1.802 z + 0.253 z^2} \,,\\
F_T^{\pi {\rm II}}(z) & = 0.319 \frac{1+ 0.328 z + 0.00274 z^2}{1 + 1.872 z + 0.350 z^2}\,.
\end{align}
\end{subequations}
They are built with the understanding that SCI $\pi$ form factors are hard, owing to the presence of $F_\pi\neq 0$ in the pion Bethe-Salpeter amplitude; see, \emph{e.g}., Ref.\,\cite{Chen:2012txa}.
}

\section{Rho-Meson Tensor Form Factors}
\label{sec4}
Analogous to the $\pi$ discussion, the $\rho$-meson tensor vertex can be expressed as follows:
\begin{align}
\Lambda_{\alpha\beta;\mu\nu}^{\rho u} (K,Q)
& = \sum_{l=1}^5 T_{\alpha\beta;\mu\nu}^{\rho l}(K,Q) F_{Tl}^{\rho u}(Q^2)\,.
\label{rhoTdef}
\end{align}
In writing this vertex, we have exploited constraints imposed by $[\mu,\nu]$-antisymmetry, parity, charge conjugation, Bose symmetry, and the $\rho$-meson being on-shell; so, there are only five independent tensors in Eq.\,\eqref{rhoTdef}, \emph{i.e}., with ${\mathpzc P}_{\alpha\beta}^P = \delta_{\alpha\beta} - P_\alpha P_\beta/P^2$:
\begin{subequations}
\label{rhoT}
\begin{align}
T_{\alpha\beta;\mu\nu}^{\rho l}(K,Q) & =
{\mathpzc P}_{\alpha\alpha^\prime}^{P_{\rm o}} \check T_{\alpha^\prime\beta^\prime;\mu\nu}^{\rho l}(K,Q)
{\mathpzc P}_{\beta^\prime\beta}^{P_{\rm i}} \\
 \check T_{\alpha\beta;\mu\nu}^{\rho 1}(K,Q) &
 = (\delta_{\alpha\mu} \delta_{\beta\nu} - \delta_{\alpha\nu} \delta_{\beta\mu})m_\rho \,, \\
 \check T_{\alpha\beta;\mu\nu}^{\rho 2}(K,Q) & =
 \left[K_{\alpha}(K_{\mu} \delta_{\beta \nu}-K_{\nu} \delta_{\beta \mu})\right. \nonumber \\
    & \left. \quad \!\!  -K_{\beta}(K_{\mu} \delta_{\alpha \nu}-K_{\nu} \delta_{\alpha \mu})\right] /(2m_{\rho})\,, \\
 \check T_{\alpha \beta ;\mu\nu}^{\rho 3}(K,Q) & =
    \left[Q_{\alpha}(Q_{\mu} \delta_{\beta \nu}-Q_{\nu} \delta_{\beta \mu})\right. \nonumber\\
    & \left. \quad \!\!  -Q_{\beta}(Q_{\mu} \delta_{\alpha \nu}-Q_{\nu} \delta_{\alpha \mu})\right] /(2m_{\rho})\,, \\
 \check T_{\alpha \beta ;\mu\nu}^{\rho 4}(K,Q) &
 =\delta_{\alpha \beta}(K_{\mu} Q_{\nu}-K_{\nu} Q_{\mu})/m_{\rho} \,, \\
 \check T_{\alpha \beta ;\mu\nu}^{\rho 5}(K,Q) &
 = K_{\alpha} K_{\beta}(K_{\mu} Q_{\nu}-K_{\nu} Q_{\mu})/m_{\rho}^{3}\,.
\end{align}
\end{subequations}
In arriving at this structure for a spin-$1$ system, we agree with Ref.\,\cite{Cosyn:2018rdm}.

The $\rho$-meson is not a stable hadron.  It decays predominantly to $\pi \pi$.  However, the width/mass ratio is less than $0.2$.
Consequently, in an approach exploited elsewhere -- see, \emph{e.g}., Refs.\,\cite{Jarecke:2002xd, Qin:2011dd, Hilger:2015ora, Williams:2015cvx, Xu:2021mju}, one may reliably treat the $\rho$ as stable in first approximation and consider decays and final state interaction contributions therefrom as subsequent perturbations.
In this case, one can work with the SCI RL expression for the $\rho$ tensor vertex, which is as follows (again, from here, we drop the $u$-quark superscript):
\begin{align}
&  \Lambda_{\alpha \beta; \mu \nu}^{\rho}(K, Q ; \zeta)
 =2 N_{c} \operatorname{tr}_{\mathrm{D}} \int_{dq}
 \left[i \bar{\Gamma}_{\beta}^{\rho}(-P_{o}) \right. \nonumber \\
& \;\; \qquad \left. \!\! \times S(q+P_{o})
 i \Gamma_{\mu \nu}^{T}(Q)
 S(q+P_{i}) i \Gamma_{\alpha}^{\rho}(P_{i}) S(q)\right]\,,
\end{align}
where $\Gamma^\rho$ is the physically normalised $\rho$-meson Bethe-Salpeter amplitude, Eq.\,\eqref{rhodecay}.

\begin{figure}[t]
\vspace*{-1ex}

\includegraphics[width=0.45\textwidth]{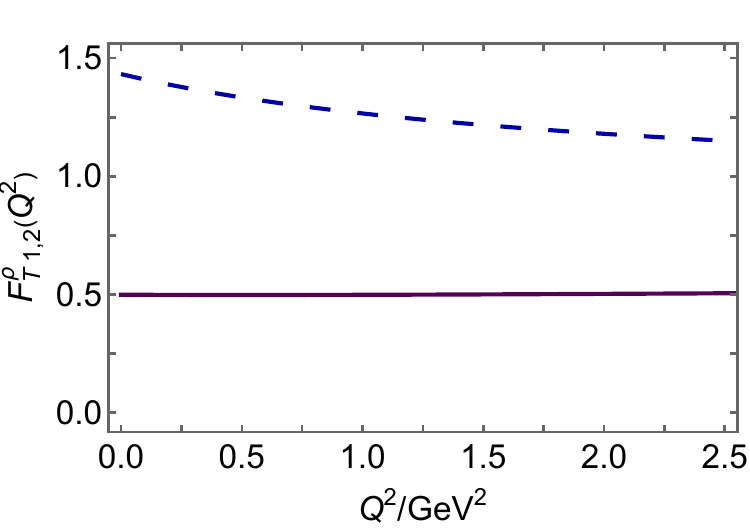}
\vspace*{-41ex}

\leftline{\hspace*{0.5em}{\large{\textsf{A}}}}

\vspace*{40ex}

\includegraphics[width=0.45\textwidth]{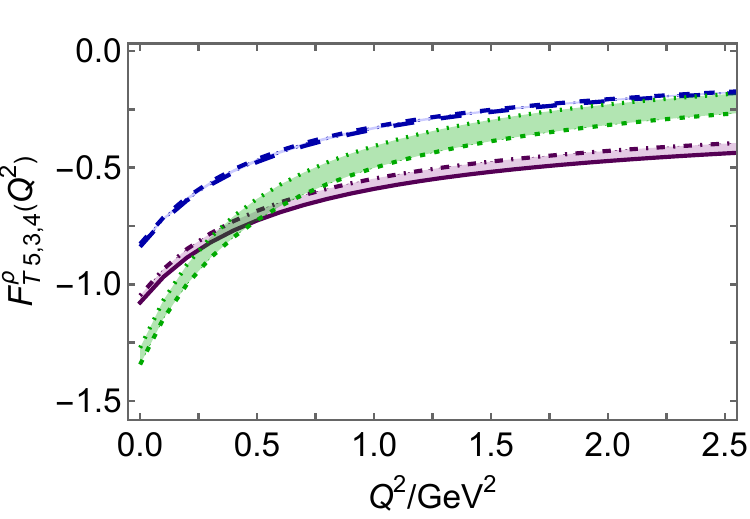}
\vspace*{-41ex}

\leftline{\hspace*{0.5em}{\large{\textsf{B}}}}

\vspace*{39ex}
\caption{\label{FigrhoT}
$u$-in-$\rho$-meson elastic tensor form factors at resolving scale $\zeta=\zeta_2$.
{\sf Panel A}.
$F_{T1}^{\rho}(Q^2)$:
solid purple.
$F_{T2}^{\rho}(Q^2)$:
long-dashed blue.
Both regularisation schemes yield identical results.
{\sf Panel B}.
$F_{T3}^{\rho}(Q^2)$:
solid purple -- Scheme I;
dot-dashed purple -- Scheme II.
$F_{T4}^{\rho}(Q^2)$:
long-dashed blue -- Scheme I;
dashed blue -- Scheme II.
$F_{T5}^{\rho}(Q^2)$:
short-dashed green -- Scheme I;
dotted green -- Scheme II.
}
\end{figure}

Explicit forms for the $\rho$ tensor form factors, obtained using both regularisation schemes, are given in \ref{ApprhoFT}.  Numerical results are depicted in Fig.\,\ref{FigrhoT}; and interpolations thereof, reliable on the domain $0\leq Q^2\leq 10\,$GeV$^2$, are presented in \ref{InterprhoT}.

In this case, the $l=1,2$ form factors are associated with tensor structures that do not involve the momentum transfer, $Q$; see Eq.\,\eqref{rhoT}.  Therefore, they receive no contribution from the ${\cal V}_2(Q^2)$ term in the quark-tensor vertex, Eq.\,\eqref{TVbsef}, and both regularisation schemes yield identical results.
Moreover, if one omits the ${\cal V}_2(Q^2)$ term from the results for $F_{T 3,4,5}^{\rho}(Q^2)$, then, for these form factors, too, identical results are produced by both schemes.  Restoring ${\cal V}_2(Q^2)$, modest differences arise.
These things are apparent in Fig.\,\ref{FigrhoT}.
For instance, $F_{T1}^\rho(Q^2)\approx\,$constant: its $Q^2=0,10\,$GeV$^2$ values differ by just 9\% because changes in the various terms that constitute the form factor almost compensate for one another; see Eqs.\,\eqref{rhoF1}, \eqref{rhoF1A} and Eqs. \eqref{rhoF1N}, \eqref{rhoj1}.
It would be interesting to see how this form factor behaves in a realistic-interaction study.

By analogy with the $\pi$, the $\rho$-meson $u$ valence-dof tensor charge is defined as follows:
$\kappa_T^\rho = F_{T1}^\rho(Q^2=0;\zeta)$.
Here, we compare the SCI prediction with an array of recent results obtained using different approaches \cite[LFWF]{Shi:2022erw}, \cite[BLFQ]{Kaur:2024iwn}, \cite[NJL]{Zhang:2024plq}, \cite[ILM]{Liu:2025fuf}:
\begin{equation}
\begin{array}{l|c|cccc}
\mbox{framework} & \mbox{herein} & \mbox{LFWF} & \mbox{BLFQ} & \mbox{NJL} & \mbox{ILM}
\\\hline
\kappa_T^\rho(\zeta_2) & 0.50 & 0.69 & 0.61 & 0.62 & 0.70
\end{array}\,.
\label{kapparho}
\end{equation}
Where a $\zeta_2$ value was not listed in the source, we used AO evolution to evolve from the given model's scale to $\zeta_2$.
A simple average of the results in Eq.\,\eqref{kapparho} yields $\kappa_T^\rho(\zeta_2)=0.62(8)$, \emph{i.e}.,  a value roughly $3$-times that of the pion and 80\% of the $u$-in-proton value.

One can again define a resolving-scale-independent tensor charge radius for each $\rho$ form factor:
\begin{equation}
(r_{Tl}^\rho)^2 = -6 \frac{d}{dQ^2} \ln F_{Tl}^\rho(Q^2) \bigg|_{Q^2=0}\,.
\label{radiiD}
\end{equation}
We list SCI predictions for $F_{Tl}^\rho(0)$ and this radius in Table~\ref{RhoTensor}\,A.
Evidently, the $l=1,2$ SCI $\rho$-meson tensor form factors are harder (more pointlike) than that of the pion, whereas the $l=3,4,5$ SCI form factors are similar in stiffness.

\begin{table}[t]
\caption{ \label{RhoTensor}
{\sf Panel A}.
Calculated $Q^2=0$ values for $\rho$-meson $u$ valence-dof tensor charge form factors at resolving scale $\zeta=\zeta_2$ and associated $\zeta$-independent tensor charge radii, with the latter measured against the analogous pion value; see Table~\ref{TensorCharge}\,A.
%
%
{\sf Panel B}.
Analogous quantities for the isovector-axialvector diquark.
}
\begin{tabular*}
{\hsize}
{
r@{\extracolsep{0ptplus1fil}}
|l@{\extracolsep{0ptplus1fil}}
l@{\extracolsep{0ptplus1fil}}
l@{\extracolsep{0ptplus1fil}}
l@{\extracolsep{0ptplus1fil}}
}\hline\hline
\centering
{\sf A} \ldots\ $l\ $ & $F_{T}^{\rho {\rm I}}\ $ & $F_{T}^{\rho {\rm II}}\ $
& $r_T^{\rho {\rm I}}/r_T^{\pi {\rm I}}\ $ & $r_T^{\rho {\rm II}}/r_T^{\pi {\rm II}}\ $ \\\hline
$1\ $ & $\phantom{-}0.50\ $ & $\phantom{-}0.50\ $ &$0.088\ $  & $0.083\ $ \\
$2\ $ & $\phantom{-}1.43\ $ & $\phantom{-}1.43\ $ & $0.35\ $ & $0.33\ $ \\
$3\ $ & $-1.08\ $ & $ -1.05\ $ & $0.92\ $ & $0.91\ $\\
$4\ $ & $-0.84\ $ & $-0.83\ $ & $1.06\ $ & $1.02\ $ \\
$5\ $ & $-1.34\ $ & $-1.27\ $ & $1.12\ $ & $1.11\ $ \\
\hline\hline
{\sf B} \ldots\ $l\ $ & $F_{T}^{1^+ {\rm I}}\ $ & $F_{T}^{1^+ {\rm II}}\ $
& $r_T^{1^+ {\rm I}}/r_T^{\pi {\rm I}}\ $ & $r_T^{1^+ {\rm II}}/r_T^{\pi {\rm II}}\ $ \\\hline
$1\ $ & $\phantom{-}0.44\ $ & $\phantom{-}0.44\ $ &$0.31\ $  & $0.29\ $ \\
$2\ $ & $\phantom{-}1.13\ $ & $\phantom{-}1.13\ $ & $0.43\ $ & $0.41\ $ \\
$3\ $ & $-1.25\ $ & $ -1.29\ $ & $1.02\ $ & $1.00\ $\\
$4\ $ & $-0.89\ $ & $-0.92\ $ & $1.07\ $ & $1.04\ $ \\
$5\ $ & $-1.78\ $ & $-1.84\ $ & $1.14\ $ & $1.12\ $ \\
\hline\hline
\end{tabular*}
\end{table}

The source of this $F_{T 1,2}^{\rho}$ \emph{cf}.\ $F_{T 3,4,5}^{\rho}$ difference is evident upon consideration of Eq.\,\eqref{TensorAnswer} and the formulae in \ref{ApprhoFT}.
As noted above, the $l=1,2$ form factors receive no contribution from ${\cal V}_2(Q^2)$; hence, they do not communicate with the $\rho$-meson pole in the quark-tensor vertex.
On the other hand, ${\cal V}_2(Q^2)$ contributions are present in the $l=3,4,5$ form factors.
Improving upon the SCI treatment, introducing a momentum-dependent quark + antiquark interaction, one would find a tensor-meson pole in the ${\cal V}_1(Q^2)$ component of $F_{T 1,2}^\rho(Q^2)$.
This would make these form factors somewhat softer; but since the mass-squared of the lightest tensor meson is $\approx 2.7 m_\rho^2$, one may expect them to still be harder than $F_{T 3,4,5}^\rho(Q^2)$.
Notably, the coupling to ${\cal V}_2$ also plays a role in making the form factors $F_{T 3,4,5}^\rho(Q^2)$ negative definite.

\begin{figure}[t]
\includegraphics[width=0.45\textwidth]{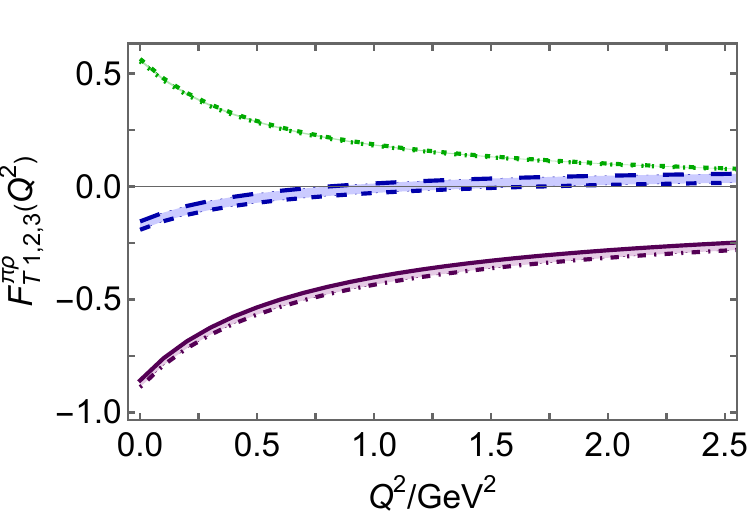}
\caption{\label{FigpirhoT}
SCI results for $\pi\leftrightarrow\rho$ tensor-induced $u$-quark transition form factors at $\zeta=\zeta_2$.
Legend.
$F_{T1}^{\pi\rho}(Q^2)$:
solid purple -- Scheme I;
dot-dashed purple -- Scheme II.
$F_{T2}^{\pi\rho}(Q^2)$:
long-dashed blue -- Scheme I;
dashed blue -- Scheme II.
$F_{T3}^{\pi\rho}(Q^2)$:
short-dashed green -- Scheme I;
dotted green -- Scheme II.
}
\end{figure}

\section{Pi-Rho Transition Tensor Form Factors}
\label{sec5}
We are now in a position to consider the $\pi \leftrightarrow \rho$ ($u$-in-$\pi\leftrightarrow u$-in-$\rho$) transitions induced by a tensor probe, which are described by the following vertex:
\begin{align}
\Lambda_{\alpha;\mu\nu}^{\pi\rho u}(K,Q) &
= \sum_{l=1}^{3} T_{\alpha;\mu\nu}^{\pi\rho l}F_{T l}^{\pi\rho u}(Q^2)\,,
\end{align}
with, in this case, $P_{\rm i}^2=-m_\rho^2$, $P_{\rm o}^2=-m_\pi^2$, $K\cdot Q = (1/2)(m_\pi^2-m_\rho^2)$; and
{\allowdisplaybreaks
\begin{subequations}
\begin{align}
T_{\alpha; \mu \nu}^{\pi\rho 1}(K, Q)
&=\epsilon_{\mu \nu \gamma \delta} {\mathpzc P}_{\alpha \gamma}^{P_{\rm i}} K_{\delta} \, , \\
T_{\alpha; \mu \nu}^{\pi\rho 2}(K, Q) &=\epsilon_{\mu \nu \gamma \delta} {\mathpzc P}_{\alpha \gamma}^{P_{\rm i}} Q_{\delta} \, , \\
T_{\alpha ; \mu \nu}^{\pi\rho 3}(K, Q) &= \epsilon_{\mu \nu \gamma \delta} P_{{\rm i}\gamma} P_{{\rm o}\delta} {\mathpzc P}_{\alpha \lambda}^{P_{\rm i}} P_{{\rm o}\lambda} / m_{\rho}^{2} \, .
\end{align}
\end{subequations}
The SCI RL matrix element for this vertex is as follows (again, hereafter, we drop the $u$-quark superscript):
\begin{align}
\Lambda_{\alpha ; \mu \nu}^{\pi \rho}& (K, Q) = 2 N_{c} \operatorname{tr}_{\mathrm{D}} \int_{dq}\left[i\bar{\Gamma}_{\pi}(-P_{\rm o}) S(q+P_{\rm i})  \right. \nonumber \\
& \left. \times  i \Gamma_{\mu \nu}^{T}(Q )
S(q-Q+P_{\rm i}) i \Gamma_{\alpha}^{\rho}(P_{\rm i}) S(q-Q)\right].
\end{align}
}

Explicit forms for the $\pi\leftrightarrow\rho$ tensor form factors, obtained using both regularisation schemes, are given in \ref{Apppirho}.
The numerical results are displayed in Fig.\,\ref{FigpirhoT}.
Evidently the transition form factors exhibit limited sensitivity to the regularisation scheme.
Interpolations of these results, reliable on the domain $0\leq Q^2\leq 10\,$GeV$^2$, are presented in \ref{ApppirhoI}.

%

Selected $Q^2$ values of $\pi\leftrightarrow \rho$ tensor transition form factors are listed in Table~\ref{piRhoTensor}\,A.
The radii are defined analogously to, \emph{e.g}., Eq.\,\eqref{radiiD}.
%
One observes that Scheme~II delivers somewhat stiffer form factors; otherwise, because each receives a $\mathcal{V}_{2}(Q^{2})$ contribution, the radii are somewhat similar to those of the $l=3,4,5$ $\rho$-meson tensor form factors.
\emph{N.B}. One should bear in mind that a small $Q^2=0$ value for a form factor may distort the associated radius' magnitude.


\begin{table}[t]
\caption{ \label{piRhoTensor}
{\sf Panel A}.
Calculated $Q^2=0$ values for
$\pi\leftrightarrow\rho$ tensor-induced $u$-valence dof quark transition form factors at $\zeta=\zeta_2$ and associated $\zeta$-independent tensor charge radii, with the latter measured against the analogous pion value; see Table~\ref{TensorCharge}\,A.
%
%
{\sf Panel B}. Analogous quantities for the $0^+ \leftrightarrow 1^+$ transition.
The ``$i$'' means the charge-radius-squared is negative; in this case because
$F_{T2}^{0^+ 1^+}(0)> 0$ and this form factor grows slowly with $Q^2$.
}
\begin{tabular*}
{\hsize}
{
r@{\extracolsep{0ptplus1fil}}
|l@{\extracolsep{0ptplus1fil}}
l@{\extracolsep{0ptplus1fil}}
l@{\extracolsep{0ptplus1fil}}
l@{\extracolsep{0ptplus1fil}}
}\hline\hline
\centering
{\sf A} \ldots\ $l\ $ & $F_{T}^{\pi\rho {\rm I}}\ $ & $F_{T}^{\pi\rho {\rm II}}\ $
& $r_T^{\pi\rho {\rm I}}/r_T^{\pi {\rm I}}\ $ & $r_T^{\pi\rho {\rm II}}/r_T^{\pi {\rm II}}\ $ \\\hline
$1\ $ & $-0.86\ $ & $-0.89\ $ &$0.96\ $  & $0.89\ $ \\
$2\ $ & $-0.16\ $ & $-0.19\ $ & $1.45\ $ & $1.24 \ $ \\
$3\ $ & $\phantom{-}0.57\ $ & $ \phantom{-}0.55\ $ & $1.13\ $ & $1.08\ $\\
\hline\hline
{\sf B} \ldots\ $l\ $ & $F_{T}^{0^+ 1^+ {\rm I}}\ $ & $F_{T}^{0^+ 1^+ {\rm II}}\ $
& $r_T^{0^+ 1^+ {\rm I}}/r_T^{\pi {\rm I}}\ $ & $r_T^{0^+ 1^+ {\rm II}}/r_T^{\pi {\rm II}}\ $ \\\hline
$1\ $ & $-0.78\ $ & $-0.77\ $                                    &$0.97\ $  & $0.92\ $ \\
$2\ $ & $\phantom{-}0.016\ $ & $\phantom{-}0.024\ $ & $2.42 i\ $ & $1.47 i \ $ \\
$3\ $ & $\phantom{-}0.66\ $ & $ \phantom{-}0.66\ $ & $1.14\ $ & $1.08\ $\\
\hline\hline
\end{tabular*}
\end{table}

\section{Tensor form factors involving  scalar and axialvector diquarks}
\label{sec6}
With the above $\pi$ and $\rho$ tensor form factors in hand, one can capitalise on the meson-diquark correspondence discussed in Ref.\,\cite{Cahill:1987qr} and exploited, \emph{e.g}., in Refs.\,\cite{Roberts:2011cf, Roberts:2011wy}, to obtain analogous form factors for the isoscalar-scalar ($0^+$) and isovector-vector ($1^+$) quark + quark (diquark) correlations, which have long been argued to play a significant role in baryon structure \cite{Barabanov:2020jvn, Yin:2021uom, Ding:2022ows, Achenbach:2025kfx}.
Such results can be useful in developing predictions for baryon tensor charges.

To obtain the diquark results from the formulae already presented herein, one need only complete the following steps \cite{Roberts:2011cf, Roberts:2011wy}.
\smallskip

\noindent (\emph{i}) Solve the scalar and axialvector diquark Bethe-Sal\-peter equations described in Ref.\,\cite[Sect.\,2]{Roberts:2011cf}, using the SCI formulation recapitulated in \ref{AppendixSCI}, to obtain the masses and amplitudes for these correlations.
This procedure yields:
\begin{equation}
\begin{array}{ccc|cc}
m_{0^+}/{\rm GeV} & E_{0^+} & F_{0^+}  & m_{1^+}/{\rm GeV} & E_{1^+} \\\hline
0.78 & 2.71 & 0.31 & 0.92 & 1.19
\end{array}\,.
\end{equation}
The meson-diquark correspondence entails
\begin{equation}
E_{1^+}^2 = E_\rho^2 \frac{3}{2} \frac{K_{\gamma}^\prime(-m_\rho^2)}{K_{\gamma}^\prime (-m_{1^+}^2)}\,,
\end{equation}
where here $[\cdot]^\prime$ denotes differentiation of the function and evaluation of the derivative at the indicated argument.

\smallskip

\noindent (\emph{ii}) With the exception of the quark-tensor-vertex form factors, $\{\mathcal{V}_{j}(Q^{2})\}$, which are tied to the $\rho$-meson, make these replacements in the meson formulae:
{\allowdisplaybreaks
\begin{subequations}
\begin{align}
F_{T}^{0^+}(Q^2) & = 
F_{T}^{\pi}(Q^2)\big|_{m_\pi \to m_{0^+}}^{(E_\pi,F_\pi) \to \surd\tfrac{2}{3} (E_{0^+},F_{0^+})} \,, \\
F_{T j}^{1^+}(Q^2) & = 
F_{T j}^{\rho}(Q^2)\big|_{m_\rho \to m_{1^+}}^{E_\rho \to \surd\tfrac{2}{3} E_{1^+}} \label{avqqFF}\\
F_{Tj}^{0^+ 1^+}(Q^2) & = 
F_{T j}^{\pi\rho}(Q^2)
\big|_{m_\pi \to m_{0^+}, m_\rho \to m_{1^+}}^{(E_\pi,F_\pi,E_\rho) \to \surd\tfrac{2}{3} (E_{0^+},F_{0^+},E_{1^+})}\,,
\end{align}
\end{subequations}
along with kindred kinematic changes in the associated tensor structures.
We note that there are three axialvector diquark correlations, \emph{viz}.\ $\{uu\}$, $\{ud\}$, $\{dd\}$ symmetric flavour combinations, which are mass degenerate.
In the tensor form factor case, unlike electromagnetic form factors, the $u$ and $d$ quarks have the same ``charge'', Eq.\,\eqref{zHdeltaq}, so one need not distinguish between these three members of the isotriplet.
}

Interpolations of the diquark form factors thereby obtained, reliable on $0\leq Q^2 \leq 10\,$GeV$^2$, are listed in \ref{AppE}.

Selected $Q^2$ values of tensor form factor properties are listed in Tables~\ref{TensorCharge}\,B, \ref{RhoTensor}\,B, \ref{piRhoTensor}\,B.
One sees therein that, apart from the mass-dependent inflation of the $0^+$ tensor charge with respect to that of the pion, tensor form factors involving $0^+$, $1^+$ diquarks are qualitatively and semiquantitatively similar to those of their $\pi$, $\rho$ counterparts.
It will be observed that the $F_{T2}^{0^+ 1^+}$ charge-radius-squared is negative because $F_{T2}^{0^+ 1^+}(0)> 0$ and this form factor grows slowly with $Q^2$.
A slight increase in $m_{1^+}$ makes $F_{T2}^{0^+ 1^+}(0)< 0$, restoring a positive charge-radius-squared.
Effectively, this transition form factor can be treated as zero.
The same is almost true of its $\pi \leftrightarrow \rho$ analogue; see Fig.\,\ref{FigpirhoT}.

\section{Summary and Perspective}
\label{epilogue}
We employed a symmetry-preserving treatment of a vector\,$\otimes$\,vector contact interaction (SCI) to deliver predictions for an array of tensor charges and form factors, elastic and transition, involving $\pi$- and $\rho$-mesons and their scalar and axialvector diquark partners.
Two distinct SCI regularisation schemes were used and the results compared.  In most cases, the differences were immaterial, adding to confidence in the robustness and value of the SCI results.

The SCI predictions for the pion tensor charge and form factor [Sect.\,\ref{sec2}] are in fair agreement with existing results obtained via numerical simulations of lattice-regularised quantum chromodynamics (QCD), with the pion's tensor radius being approximately identical to its charge radius.
This confluence highlights the utility of SCI analyses: it is a simple, insightful framework delivering sound predictions.

The $\rho$-meson has five independent tensor form factors [Sect.\,\ref{sec4}], one of which is directly related to the tensor charge that is commonly extracted via analyses of $\rho$-meson generalised parton distributions.  Our SCI analysis predicts $\kappa_T^\rho = 0.50$ at resolving scale $\zeta = \zeta_2 := 2\,$GeV.  Combining that result with those obtained using a raft of other methods, one arrives at the estimate $\kappa_T^\rho(\zeta_2) = 0.62(8)$ for the $u$ quark in the $\rho$-meson.  This is roughly 80\% of the in-proton value.
The $\rho$-meson elastic tensor form factors exhibit little sensitivity to the regularisation scheme.

Alike with the photon, a tensor probe can induce $\pi\leftrightarrow\rho$ transitions.  There are three associated form factors [Sect.\,\ref{sec5}].  They exhibit practically no sensitivity to the regularisation scheme.

Exploiting a widely-used meson-diquark correspondence, tensor form factors involving isoscalar-scalar and isovector-axialvector diquark correlations were obtained directly from the $\pi$- and $\rho$-meson tensor form factors [Sect.\,\ref{sec6}].  Apart from a kinematic enhancement of the $0^+$ diquark tensor charge compared with that of the pion, tensor form factors involving the diquarks are qualitatively and semiquantitatively similar to those of their $\pi$, $\rho$ analogues.  These results can be useful in future analyses of baryon tensor charges within the modern quark + diquark picture of baryon structure.

In addition to being interesting in themselves, the SCI predictions herein can serve as baselines for future studies with a closer connection to QCD, \emph{e.g}., those which use a quark + antiquark interaction with realistic momentum dependence.
Such an analysis is already underway.
As already noted, they can also be useful as inputs for analyses of more complex systems, like the nucleon and its tensor charges.  In this connection, our preliminary analysis agrees with earlier studies, which suggest that existing phenomenological inferences of the $u$-quark-in-proton tensor charge underestimate the true value by as much as a factor of two [Sect.\,\ref{QTV}].  This work is ongoing.

\begin{acknowledgements}
We are grateful for constructive comments from P.~Cheng and Y.~Yang.
Work supported by:
National Natural Science Foundation of China, under grant nos.\ 12135007, 12247103.
\end{acknowledgements}

\appendix

\section{SCI}
\label{AppendixSCI}
The SCI is described in many sources.
Nevertheless, here, for internal completeness, we reproduce and augment material from Ref.\,\cite[Appendix~A]{Cheng:2026nud}.

\subsection{Special functions}
\label{SpecialF}
Functions of the following type arise in SCI bound-state equations:
\begin{align}
%
%
%
n !\, \overline{\cal C}^{\rm iu}_n(\omega) & = \Gamma(n-1,\omega \tau_{\textrm{uv}}^{2}) - \Gamma(n-1,\omega \tau_{\textrm{ir}}^{2})\,,
\label{eq:Cn}
\end{align}
where $\tau_{\rm ir}$, $\tau_{\rm uv}$ are SCI parameters, ${\cal C}^{\rm iu}_n(\omega)=\omega \overline{\cal C}^{\rm iu}_n(\omega)$, $n\in {\mathbb Z}^\geq$, with $\Gamma(x,y)$ being the incomplete gamma function.

{\allowdisplaybreaks
In connection with these functions, there are useful differentiation and integration rules:
\begin{subequations}
\begin{align}
\label{DiffCbar}
{\cal C}_{n}^{\mathrm{iu}}(\omega)& =(-1)^{n} \frac{\omega^{n}}{n!} \frac{d^{n}}{d \omega^{n}} {\cal C}_{0}^{\mathrm{iu}}(\omega)\,,\\
%
\label{IntCbar}
\int_0^\infty dy \, (n+1)! & \frac{\overline{\cal C}^{\rm iu}_{n+1}(y+y_0)}{(y+y_0)^{n}}
= n! \frac{\overline{\cal C}^{\rm iu}_n(y_0)}{y_0^{n-1}}\,.
\end{align}
\end{subequations}}

In order to ensure that vector and axialvector Ward-Green-Takahashi identities are satisfied, the following identity is enforced
\begin{equation}
\label{C0C1}
    0=\int_{0}^{1} d \alpha\left[\mathcal{C}_{0}^{\mathrm{iu}}(\omega(\alpha, P^{2}))+\mathcal{C}_{1}^{\mathrm{iu}}(\omega(\alpha, P^{2}))\right],
\end{equation}
where
\begin{equation}
\omega(\alpha, P^{2})=m_M^{2}+\alpha(1-\alpha)P^{2}\,,
\end{equation}
with $m_M$ the meson mass.
This guarantees the absence of logarithmic and quadratic divergences in relevant four-dimensional integrals.

\subsection{SCI background}
\label{SCIback}
Using RL truncation, the quark + antiquark scattering kernel can be written:
\begin{align}
\label{KDinteraction}
\mathscr{K}_{\alpha_1\alpha_1',\alpha_2\alpha_2'}  & = \tilde{\mathpzc G}(k^2) {\mathpzc P}^k_{\mu\nu} [i\gamma_\mu]_{\alpha_1\alpha_1'} [i\gamma_\nu]_{\alpha_2\alpha_2'}\,,
\end{align}
where $k = p_1-p_1^\prime = p_2^\prime -p_2$, with $p_{1,2}$, $p_{1,2}^\prime$ being the initial and final momenta, respectively, of the scatterers, and $k^2{\mathpzc P}_{\mu\nu}^k = k^2\delta_{\mu\nu} - k_\mu k_\nu$.

The key piece in Eq.\,\eqref{KDinteraction} is $\tilde{\mathpzc G}$.
Analyses of QCD gauge sector dynamics \cite{Gao:2017uox, Cui:2019dwv} have revealed that a gluon mass-scale emerges in QCD; hence, $\tilde{\mathpzc G}$ is nonzero and finite at infrared momenta:
\begin{align}
\label{SimpG}
\tilde{\mathpzc G}(k^2) & \stackrel{k^2 \simeq 0}{=} \frac{4\pi \alpha_{\rm IR}}{m_G^2}\,,
\end{align}
with \cite{Cui:2019dwv, Deur:2023dzc, Brodsky:2024zev}: $m_G \approx 0.5\,$GeV, $\alpha_{\rm IR} \approx \pi$.
In our analyses, we keep the QCD value of $m_G$.
In addition, capitalising on the fact that a SCI does not support relative momentum between valence dof in a bound state, the tensor in Eq.\,\eqref{KDinteraction} can be simplified, \emph{viz}.\
\begin{align}
\label{KCI}
\mathscr{K}_{\alpha_1\alpha_1',\alpha_2\alpha_2'}^{\rm CI}  & = \frac{4\pi \alpha_{\rm IR}}{m_G^2}
 [i\gamma_\mu]_{\alpha_1\alpha_1'} [i\gamma_\mu]_{\alpha_2\alpha_2'}\,.
\end{align}

As detailed below, confinement is implemented by including an infrared mass scale, $\Lambda_{\rm ir}$, when solving all equations that appear in bound-state problems \cite{Ebert:1996vx}: $\Lambda_{\rm ir}>0$
ensures the absence of quark + antiquark production thresholds \cite{Krein:1990sf}.
Typically \cite{Gutierrez-Guerrero:2010waf}: $\Lambda_{\rm ir} = 0.24\,$GeV $=1/[0.82\,{\rm fm}]$, \emph{i.e}., a confinement length that roughly matches the proton size \cite{Cui:2022fyr}.

SCI integrals require ultraviolet regularisation, too.
This breaks the link between ultraviolet and infrared scales that characterises QCD.
So, the associated ultraviolet mass-scale, $\Lambda_{\rm uv}$, becomes a physical parameter.
It may fairly be interpreted as an upper bound on the domain upon which amplitudes describing the associated bound-states are practically momentum-independent.

Two methods for regulating SCI ultraviolet divergences have been used:
the first was discussed in Ref.\,\cite{Gutierrez-Guerrero:2010waf}; and another was used in Ref.\,\cite{Xing:2022jtt}.
Herein, we will compare results obtained with both.
To begin, we describe Scheme~I results \cite{Gutierrez-Guerrero:2010waf}.

\begin{table}[t]
\caption{\label{Tab:DressedQuarks}
Block 1.
SCI inputs: coupling, $\alpha_{\rm IR}$, ultraviolet cutoff, $\Lambda_{\rm uv}$, and hadron-scale current-quark masses, $m_{u,s}$, which enable a good description of flavour-nonsinglet pseudoscalar- and $\rho$-meson properties.
As is standard, $m_G=0.5\,$GeV, $\Lambda_{\rm ir} = 0.24\,$GeV.
Block 2. Calculated results: dressed-quark masses, $M_{u,s}$; meson masses, $m_{\pi,\rho,K}$; meson leptonic decay constants $f_{\pi,\rho,K}$.
With our normalisation, empirical values for the masses and
decay constants are
$0.14$, $0.78$, $0.49$ and
$0.092$, $0.15$, $0.11$, respectively.
%
%
%
(We assume isospin symmetry and list dimensioned quantities in GeV.)}
\begin{center}
\begin{tabular*}
{\hsize}
{
l@{\extracolsep{0ptplus1fil}}
|c@{\extracolsep{0ptplus1fil}}
c@{\extracolsep{0ptplus1fil}}
c@{\extracolsep{0ptplus1fil}}
c@{\extracolsep{0ptplus1fil}}
|c@{\extracolsep{0ptplus1fil}}
c@{\extracolsep{0ptplus1fil}}
c@{\extracolsep{0ptplus1fil}}
c@{\extracolsep{0ptplus1fil}}
c@{\extracolsep{0ptplus1fil}}
c@{\extracolsep{0ptplus1fil}}}\hline
$\alpha_{\rm IR}\ $ & $\alpha_{\rm IR}/\pi\ $ & $\Lambda_{\rm uv}$ & $m_u$ & $m_s$  &   $M_u$ & $M_s\ $ & $m_{\pi,\rho,K}$ & $f_{\pi,\rho,K}$ \\\hline
$\ell=u/d_\pi\ $  & $0.36\ $ & $0.91\ $ & $0.007\ $ & & $0.37\ $ & &  $0.14\phantom{5}\ $ & $0.10\ $  \\\hline
$\ell=u/d_\rho\ $  & $0.76\ $ & $0.91\ $ & $0.007\ $ & & $0.37\ $ & &  $0.775\ $ & $0.15\ $  \\\hline
$s_K$ & $0.33\ $ & $0.94\ $ & $0.007\ $ & $0.16\ $& $0.37\ $ & $0.53\ $ & $0.50\phantom{5}\ $ & $0.11\ $  \\\hline
\end{tabular*}
\end{center}
\end{table}

\subsection{Gap equation}
\label{GapEq}
{\allowdisplaybreaks
Consider the SCI gap equation for an $f$-flavour quark:
\begin{align}
\label{GapEqn}
S_f^{-1}(p)  & = i\gamma\cdot p +m_f \nonumber \\
& \quad + \frac{16 \pi}{3} \frac{\alpha_{\rm IR}}{m_G^2}
\int \frac{d^4q}{(2\pi)^4} \gamma_\mu S_f(q) \gamma_\mu\,,
\end{align}
where $m_f$ is the associated quark current-mass.  Although only mass-degenerate light quarks are considered herein, to provide additional context, Table~\ref{Tab:DressedQuarks} also lists kaon-related results.
Using a Poincar\'e-invariant regularisation, the solution has the form:
\begin{equation}
\label{genS}
S_f^{-1}(p) = i \gamma\cdot p + M_f\,,
\end{equation}
where the SCI dressed-quark's momentum-independent dynamically generated mass, $M_f$  is obtained by solving:
\begin{equation}
M_f = m_f + M_f\frac{4\alpha_{\rm IR}}{3\pi m_G^2}\,\,{\cal C}_0^{\rm iu}(M_f^2)\,.
\label{gapactual}
\end{equation}
}

In order to arrive at Eq.\,\eqref{gapactual}, we have implemented the following regularisation steps
($\tau_{\rm uv}^2=1/\Lambda_{\textrm{uv}}^{2}$, $\tau_{\rm ir}^2=1/\Lambda_{\textrm{ir}}^{2}$):
\begin{subequations}
\label{SCIReg1}
\begin{align}
\int_0^\infty ds\, s \frac{1}{s+\sigma}
& = \int_0^\infty ds\, s \int_0^\infty d\tau {\rm e}^{-\tau[s+\sigma]} \\
& \to \int_0^\infty ds\, s \int_{\tau_{\rm uv}}^{\tau_{\rm ir}} d\tau {\rm e}^{-\tau[s+\sigma]} \\
& =: {\cal C}^{\rm iu}_0(\sigma)\,.
\end{align}
\end{subequations}
A nonzero value of $\Lambda_{\rm ir}$ implements confinement by eliminating quark production thresholds
\cite[Sect.\,5]{Ding:2022ows};
and $\tau_{\rm uv} \neq 0$ is required to produce a finite result for the integral; thus, the value of $\Lambda_{\rm uv}$ becomes a dynamical scale in the SCI; hence, a part of its definition: it sets the physical scale for all mass-dimensioned quantities and, qualitatively, an upper bound on the domain of SCI validity.

The identities in \ref{SpecialF} can be used to simplify other integrals that arise in the SCI treatment of meson properties using the Ref.\,\cite{Gutierrez-Guerrero:2010waf} scheme.

\subsection{Meson Bethe-Salpeter amplitudes}
\label{MesonBSA}
\subsubsection{$\pi$-meson}
The pion emerges as a quark +anti\-quark bound-state, whose structure is described by a Bethe-Salpeter amplitude.  In the SCI, that amplitude takes the following form ($M=M_u$):
\begin{align}
\Gamma_{\pi}(P) = \gamma_5 \left[ i E_{\pi}(P) + \frac{\gamma\cdot P }{2M}F_{\pi}(P)\right]\,,
\label{PSBSAA}
\end{align}
where
$P$ is the pion total momentum, $P^2 = -m_{\pi}^2$.
As stressed elsewhere \cite{Gutierrez-Guerrero:2010waf, Chen:2012txa}, the axialvector Ward-Green-Takahashi identity is violated if one omits the $\gamma\cdot P F_{\pi}(P)$ term.
Notwithstanding, this omission is typical in Nambu--Jona-Lasinio model treatments of hadron properties.

The pion bound-state amplitude and $m_{\pi}^2$ are obtained by solving the SCI Bethe-Salpeter equation $(t_+ = t+P)$, $S_u=S=S_d$:
\begin{align}
\Gamma_{\pi}(P)  & =  - \frac{16 \pi}{3} \frac{\alpha_{\rm IR}}{m_G^2}
\int \! \frac{d^4t}{(2\pi)^4} \gamma_\mu S(t_+) \Gamma_{\pi}(P)S(t) \gamma_\mu \,.
\label{LBSEI}
\end{align}
Completing standard spinor projections, one arrives at the following matrix equation:
\begin{equation}
\label{bsefinalE}
\left[
\begin{array}{c}
E_{\pi}(P)\\
F_{\pi}(P)
\end{array}
\right]
= \frac{4 \alpha_{\rm IR}}{3\pi m_G^2}
\left[
\begin{array}{cc}
{\cal K}_{EE}^{\pi} & {\cal K}_{EF}^{\pi} \\
{\cal K}_{FE}^{\pi} & {\cal K}_{FF}^{\pi}
\end{array}\right]
\left[\begin{array}{c}
E_{\pi}(P)\\
F_{\pi}(P)
\end{array}
\right],
\end{equation}
with ($\check \alpha = 1-\alpha$)
{\allowdisplaybreaks
\begin{subequations}
\label{fgKernel}
\begin{align}
{\cal K}_{EE}^{\pi} &=
\int_0^1d\alpha \bigg[
{\cal C}_0^{\rm iu}(\omega( \alpha, P^2))  -2 \alpha\check\alpha P^2 \overline{\cal C}^{\rm iu}_1(\omega(\alpha, P^2))\bigg],\\
{\cal K}_{EF}^{\pi} &= P^2 \int_0^1d\alpha\,
\overline{\cal C}^{\rm iu}_1(\omega(\alpha, P^2)),\\
{\cal K}_{FE}^{\pi} &= \frac{M^2}{2 P^2}{\cal K}_{EF}^{\pi},\\
{\cal K}_{FF}^{\pi} &= - 2 {\cal K}_{FE}^{\pi}\,.
\end{align}
\end{subequations}}
\hspace*{-0.2\parindent}As usual, the value of $P^2=-m_{\pi}^2$ for which Eq.\,\eqref{bsefinalE} is satisfied corresponds to the bound-state mass.
Our result is listed in Table~\ref{Tab:DressedQuarks}, and the associated solution vector is the pion's Bethe-Salpeter amplitude.

When calculating observables, one must use the ca\-nonically normalised bound-state amplitude: \emph{i.e}., the amplitude obtained after rescaling such that
\begin{equation}
\label{normcan}
1=\left. \frac{d}{d P^2}\Pi_{\pi}(Z,P)\right|_{Z=P},
\end{equation}
where:
\begin{align}
\Pi_{\pi}(Z,Q) & = 6 {\rm tr}_{\rm D} \!\! \int\! \frac{d^4t}{(2\pi)^4}   \Gamma_{\pi}(-Z)
 S(t_+) \, \Gamma_{\pi}(Z)\, S(t)\,.
 \label{normcan2}
\end{align}
Implementing this, the dimensionless pion amplitude is:
\begin{align}
E_\pi & = 3.59\,, \quad F_\pi = 0.47\,.
\label{pionBSA}
\end{align}

With the canonically normalised Bethe-Salpe\-ter amplitude in hand, the pion leptonic decay constant is
\begin{align}
f_{\pi} &= \frac{N_c}{2\pi^2 M}\,
\big[ E_{\pi} {\cal K}_{FE}^{\pi} + F_{\pi}{\cal K}_{FF}^{\pi} \big]_{P^2=-m_{\pi}^2}\,. \label{ffg}
\end{align}

In calculating observables, the charge-conjugated amplitude is often necessary, \emph{e.g}.,
\begin{equation}
\bar \Gamma^\pi (P) = C [\Gamma^\pi(P)]^{\rm t} C^\dagger \,,
\label{ChargeC}
\end{equation}
where $C=\gamma_2\gamma_4$ is the charge conjugation matrix and $[\cdot]^{\rm t}$ indicates matrix transpose.

\subsubsection{$\rho$-meson}
\label{Newrho}
The SCI does not support relative momentum between bound-state valence dof; consequently, instead of involving eight independent amplitudes \cite{LlewellynSmith:1969az, Krassnigg:2010mh}, the $\rho$ Bethe-Salpeter amplitude takes the following simple form:
\begin{equation}
\Gamma_{\mu}^{\rho}(P)=\gamma_{\mu}^{T} E_{\rho}(P),
    \label{rhobsef}
\end{equation}
where $P_{\mu}\gamma_{\mu}^{\rm T}=0$ and $\gamma_{\mu}^{\rm T}=\gamma_{\mu}-P_{\mu}\gamma\cdot P/P^{2}$.
Making the replacement $\Gamma_\pi \to  \Gamma_{\mu}^{\rho}$ in Eq.\,\eqref{PSBSAA}, one obtains the explicit form of the associated Bethe-Salpeter equation \cite{Roberts:2011wy}:
\begin{equation}
1+K_{\gamma}(-m_{\rho}^{2})=0,
    \label{rhoren}
\end{equation}
with $K_{\gamma}$ given in Eq.\,\eqref{kgamma}.
The issue is that with only this single (leading) amplitude, the $\rho$ mass is 20\% too large \cite{Roberts:2011wy}: $m_\rho = 0.93\,$GeV.  (This outcome is also found in realistic interaction studies \cite{Maris:1999nt}.)

Given the potential sensitivity of tensor form factors to the mass of the hadron(s) involved -- see the discussion of Fig.\,\ref{FigpionT}, we choose to correct for this mass overestimate by using a simple expedient; namely, we increase $\alpha_{\rm IR}$ in the vector channel so that Eq.\,\eqref{rhoren} returns $m_\rho = 0.775\,$GeV.  This leads to the value in Table~\ref{Tab:DressedQuarks}\,-\,Row~2.

    \label{rhobsa}
In terms of the $\rho$-meson amplitude, the leptonic decay constant is \cite{Roberts:2011wy}
\begin{equation}
f_{\rho} =
-E_{\rho}\frac{3N_{c}m_{G}^{2}}{8\pi m_{\rho}}K_{\gamma}(-m_{\rho}^{2})\,.
    \label{rhodecay}
\end{equation}
Requiring that this formula yield the empirical value, $f_\rho = 0.153\,$GeV, one arrives at a physically normalised $\rho$ amplitude, \emph{viz}.\, $E_\rho = 1.32$.

Since the $\rho$-meson also contributes to the quark photon vertex, then these amendments also affect the pion elastic electromagnetic form factor.  Updating the work in Ref.\,\cite{Roberts:2011wy} accordingly, one finds
\begin{equation}
F_\pi(z=Q^2) =
\frac{1 + 0.33 z + 0.035 z^2}{1 + 1.67 z + 0.064 z^2}\,,
\label{Fpiem}
\end{equation}
which yields the following charge radius: $r_\pi^{\rm SCI} = 0.56\,$fm.

\subsubsection{SCI tensor vertex}
\label{TensorV}
To begin, recall that an axialvector ($J^{PC}=1^{++}$) fer\-mion + antifermion vertex, $\Gamma_{5\mu}(l;Q)$, contains, in general, twelve independent Dirac-matrix-valued terms, four of which have overlap with the pseudoscalar channel; hence, pseudoscalar ($0^{-+}$) meson poles appear in the $1^{++}$ vertex; see, \emph{e.g}., Ref.\,\cite{Maris:1997tm}.  This is merely an expression of partial conservation of the axialvector current.
The SCI axialvector vertex has only three terms because it cannot depend on relative quark-antiquark momentum \cite[Appendix\;A]{Xing:2022sor}.  Only one of the three is purely spin-one ($1^{++}$).  It is identified by imposing the subsidiary condition $Q_\mu\Gamma_{5\mu}=0$.  The other two (longitudinal terms) correspond to lower-spin $0^{-+}$ contributions.  

Against this background, consider now the SCI tensor vertex, Eq.\,\eqref{TVbsef}, repeated here for convenience: 
\begin{align}
    \Gamma_{\mu \nu}^{\rm T}(Q ) & =\mathcal{V}_{1}(Q^{2}) \sigma_{\mu \nu}
    +\mathcal{V}_{2}(Q^{2}) \frac{1}{M}[i\gamma\cdot Q,\sigma_{\mu \nu}]\nonumber \\
    & \quad +\mathcal{V}_{3}(Q^{2})\frac{1}{M^2} i\gamma\cdot Q \sigma_{\mu \nu} i\gamma\cdot Q .
\label{TVbsefN}
\end{align}
One should expect similar spin/parity-mixing features as in the axialvector vertex.

For a start, under charge conjugation, Eq.\,\eqref{ChargeC}:
\begin{equation}
    \Gamma_{\mu \nu}^{\rm T}(Q )  \to C  [  \Gamma_{\mu \nu}^{\rm T}(Q ) ]^{\rm t} C^\dagger = -     \Gamma_{\mu \nu}^{\rm T}(Q ) .
\end{equation}
Thus, a tensor vertex seeded by $\sigma_{\mu\nu}$ is an eigenvalue $(-1)$ eigenvector of the charge conjugation operator. 
This automatically eliminates any connection between such a vertex and quark model tensor mesons.

Next, note that, in general, a (massive) pure $J=2$ system is described by a symmetric rank-$2$ tensor, $T_{\mu\nu}(l,Q)$, which is divergenceless and traceless, \emph{viz}.\  
\begin{equation}
Q_\mu T_{\mu\nu}(l,Q) = 0 = T_{\mu\mu}(l,Q) \,.
\end{equation}
These two subsidiary conditions eliminate ``contamination'' by $J=1$ and $J=0$ states, respectively.  

Since the inhomogeneity in Eq.\,\eqref{TVbse} is $\sigma_{\mu\nu}$, which is antisymmetric, the vertex $\Gamma_{\mu \nu}^{\rm T}(Q)$ cannot exhibit any structure associated with a $J=2$ (tensor) meson (of any charge conjugation parity).  Indeed, in a Poincar\'e invariant quantum field theory, quark model analogues of tensor meson bound states appear as poles in inhomogeneous Bethe-Salpeter equations driven by $\{\mu,\nu\}$ symmetric inhomogeneities; see, \emph{e.g.}, Refs.\,\cite{LlewellynSmith:1969az, Krassnigg:2010mh}.
Furthermore, since any antisymmetric rank-$2$ tensor is traceless, \emph{viz}.\  $\Gamma_{\mu \mu}^{\rm T} =  0$, it cannot have any overlap with scalar mesons, either.  

At this point it is plain that $\Gamma_{\mu \nu}^{\rm T}(Q )$ can only connect with $J^{PC}=1^{+-}, 1^{--}$ channels.  
Consequently, the vertex seeded by $\sigma_{\mu\nu}$ cannot exhibit any poles associated with quark model like axialvector mesons, either.
We stress that these are general statements, not particular to SCI treatments. 

Now consider each term in Eq.\,\eqref{TVbsefN}.
%
The leading structure can be written in the following form:
\begin{subequations}
\begin{align}
\mathcal{V}_{1}(Q^{2}) \sigma_{\mu \nu} & = \mathcal{V}_{1}(Q^{2}) 
\left[ \tfrac{1}{2} \sigma_{\mu\nu}^+ + \tfrac{1}{2} \sigma_{\mu\nu}^- \right] \,, \\
\sigma_{\mu\nu}^\pm & = \sigma_{\mu\nu} \pm \gamma_5 \sigma_{\mu\nu} \,.
\end{align}
\end{subequations}
This quantity has (only) six independent terms; and under parity transformations: $\sigma_{\mu\nu}^\pm \to \gamma_5 \sigma_{\mu\nu}^\pm \gamma_5 = \pm \sigma_{\mu\nu}^\pm$.
Consequently, the first structure in $\Gamma_{\mu\nu}^T(Q)$ connects to $1^{+-}$ (three terms) and $1^{--}$ channels (remaining three terms).  The latter correspond to the standard vector meson channel. 

Turning to the $\mathcal{V}_{2}(Q^{2})$ component, one readily finds that it is proportional to $(Q_\mu \gamma_\nu - Q_\nu \gamma_\mu)$; hence, is purely $J=1^{--}$.  This fact is emphatically expressed by Eq.\,\eqref{V2}, \emph{viz}.\ $\mathcal{V}_{2}(Q^{2})$ exhibits a pole at the SCI mass of the $\rho$-meson and its nonzero residue owes to mixing with the $1^{--}$ projection of the $\mathcal{V}_{1}(Q^{2})$ component.

The remaining $\mathcal{V}_{3}(Q^{2})$ piece has both $J=1^{\pm -}$ contributions.  However, given Eq.\,\eqref{V3}, this is unimportant because the RL treatment of the SCI does not support this term.

\subsection{Regularisation Scheme II}
\label{AppendixA1Reg}
The results in \ref{GapEq}, \ref{MesonBSA}
were obtained using regularisation Scheme I.
Another, less frequently used, approach can be introduced as follows \cite{Xing:2022jtt}.  First consider:
{\allowdisplaybreaks
\begin{subequations} 
\label{CnIntegrals}
\begin{align}
{\mathsf C}_{-2\alpha}(\omega) & = \int \frac{d^4 l}{\pi^2} \frac{1}{[l^2+\omega]^{2+\alpha}}\\
& = \int \frac{d^4 l}{\pi^2} \int_0^{\infty} d\tau \frac{\tau^{1+\alpha}}{\Gamma(2+\alpha)} {\rm e}^{-\tau [l^2+\omega]}\\
& = \int_0^\infty d\tau
\frac{\tau^{\alpha-1}}{\Gamma(2+\alpha)} {\rm e}^{-\tau \omega}\,.
\end{align}
\end{subequations}
Here, each step is well defined for $\alpha>0$.
However, one encounters cases with $\alpha=-1, 0$ in SCI applications wherewith the integrals are, respectively, quadratically or logarithmically divergent.
We therefore again follow Ref.\,\cite{Ebert:1996vx} and introduce a proper-time regularisation of the integrals:
\begin{subequations}
\label{OldNew}
\begin{align}
& {\mathsf C}_{-2\alpha}^{\rm iu} (\omega)
= \int_{\tau_{\rm uv}}^{\tau_{\rm ir}} d\tau
\frac{\tau^{\alpha-1}}{\Gamma(2+\alpha)} {\rm e}^{-\tau \omega} \\
& = \frac{1}{\Gamma(2+\alpha)}  \frac{1}{\omega^\alpha}
[
\Gamma(\alpha, \tau_{\rm uv}\omega)
-\Gamma(\alpha, \tau_{\rm ir}\omega) ] \\
& = \frac{1}{\omega^\alpha} \overline{\cal C}^{\rm iu}_{\alpha+1}(\omega)\,, \label{OldRegNew}
%
\end{align}
\end{subequations}
where
$\overline{\cal C}^{\rm iu}_{\alpha+1}(\omega)$ are the regularisation functions introduced in \ref{SpecialF}.
}

Another case is
\begin{subequations}
\label{Dintegral}
\begin{align}
{\mathsf D}_{-2\alpha}&(\omega)  \delta_{\mu\nu} =
\int \frac{d^4 l}{\pi^2} \frac{l_\mu l_\nu}{[l^2+\omega]^{3+\alpha}}\\
& = \delta_{\mu\nu} \int \frac{d^4 l}{\pi^2} \frac{l^2}{4}
\int_0^\infty  d\tau \frac{\tau^{2+\alpha}}{\Gamma(3+\alpha)} {\rm e}^{-\tau [l^2+\omega]}\\
& = \delta_{\mu\nu} \frac{1}{2}
 \int_0^\infty  d\tau  \frac{\tau^{\alpha-1}}{\Gamma(3+\alpha)} {\rm e}^{-\tau \omega}\,.
\end{align}
\end{subequations}
Now introduce
\begin{subequations}
\begin{align}
 {\mathsf D}_{-2\alpha}^{\rm iu} (\omega)
& = \frac{1}{2} \int_{\tau_{\rm uv}}^{\tau_{\rm ir}} d\tau
\frac{\tau^{\alpha-1}}{\Gamma(3+\alpha)} {\rm e}^{-\tau \omega} \\
& = \frac{1}{2} \frac{\Gamma(2+\alpha)}{\Gamma(3+\alpha)}
{\mathsf C}_{-2\alpha}^{\rm iu} (\omega)\\
& = \frac{1}{2} \frac{1}{2+\alpha}
{\mathsf C}_{-2\alpha}^{\rm iu} (\omega)\,.
\end{align}
\end{subequations}

Following these procedures, one is expressing the following identity:
\begin{align}
& 4 {\mathsf D}_{-2\alpha}^{\rm iu} (\omega) =
\int d^4 l \frac{l^2}{[l^2+\omega]^{3+\alpha}} \nonumber \\
%
& \stackrel{\rm reg.}{=}
\frac{2}{2+\alpha}
\int d^4 l \frac{1}{[l^2+\omega]^{2+\alpha}} =
\frac{2}{2+\alpha}{\mathsf C}_{-2\alpha}^{\rm iu} (\omega)\,.
\label{IntegralIdentities}
\end{align}
Naturally, this is true for any value of $\alpha$ for which both integrals are finite.  By analytic continuation, regularisation defines Eq.\,\eqref{IntegralIdentities} to be true at all other $\alpha$ values.

Note that:
\begin{equation}
{\mathsf D}_{0}^{\rm iu} (\omega) =  (1/4){\mathsf C}_{0}^{\rm iu} (\omega)\,,
{\mathsf D}_{2}^{\rm iu} (\omega) =  (1/2){\mathsf C}_{2}^{\rm iu} (\omega)\,.
\end{equation}
In ensuring, \emph{e.g}., the axialvector Ward-Green-Takahashi identity, the latter relation merely suggests a different rearrangement of terms in the integrand so as to ensure Ref.\,\cite[Eq.\,(17)]{Gutierrez-Guerrero:2010waf}, \emph{viz}.
\begin{align}
\int \frac{d^4 l}{\pi^2} \frac{\tfrac{1}{2} l^2 + \omega}{[l^2 + \omega]^2} &
\to \frac{1}{2} [ {\cal C}_0^{\rm ir}(\omega) + {\cal C}_1^{\rm ir}(\omega) ] \\
&
= \int \frac{d^4 l}{\pi^2}
\left[ \frac{1}{l^2 + \omega} - \frac{1}{2} \frac{l^2}{[l^2+\omega]^2}\right] \\
& \to {\mathsf C}_{2}^{\rm ir}(\omega) - 2 {\mathsf D}_{2}^{\rm ir}(\omega) = 0 \,.
%
\end{align}
This elucidates the connection between Eq.\,\eqref{IntegralIdentities} and Eq.\,\eqref{C0C1} and serves to ensure that the axialvector Ward-Green-Takahashi identity is satisfied.

\subsection{Scheme Differences}
\label{SchemeDiff}
Here, it is worth recording the origin of mismatches between tensor form factors obtained with the two different regularisation schemes.  To that end, consider the following integral
\begin{equation}
\int_{dt} \frac{A t^2  + 4 B (u\cdot t)^2 }{(t^2 + \omega)^3}\,,
\end{equation}
which arises in the calculation of tensor form factors and is $\ln$-divergent.

Using Scheme I, Eqs.\,\eqref{eq:Cn}, \eqref{DiffCbar}, one finds
\begin{align}
\int_{dt}& \frac{A t^2  + 4 B (u\cdot t)^2 }{(t^2 + \omega)^3} \nonumber \\
&  =
 (A + B u^2) \big[
\bar{\mathcal{C}}^{\mathrm{iu}}_{1}(\omega) -\bar{\mathcal{C}}^{\mathrm{iu}}_{2}(\omega)  \big]\,,
\end{align}
whereas, with Scheme II
\begin{align}
\int_{dt}& \frac{A t^2  + 4 B (u\cdot t)^2 }{(t^2 + \omega)^3}  =
 (A + B u^2) \mathsf{C}^{\mathrm{iu}}_{0}(\omega)
- A \omega \mathsf{C}^{\mathrm{iu}}_{-2} \nonumber \\
& =
 (A+B u^2) \bar{\mathcal{C}}^{\mathrm{iu}}_{1}(\omega)
- A \bar{\mathcal{C}}^{\mathrm{iu}}_{2}(\omega)
\end{align}
where we have used Eq.\,\eqref{OldRegNew}.  Evidently, the two regularisation schemes introduce distinct definitions of the regularised integral, with the difference being
\begin{equation}
\mbox{Scheme I - Scheme II} = - B u^2\bar{\mathcal{C}}^{\mathrm{iu}}_{2}(\omega) \,.
\end{equation}

\section{Pion Tensor Form Factor}
\label{ApppionFT}
One can write the pion's dimensionless tensor form factor in the following form:
\begin{equation}
\begin{aligned}
F_{T}^{\pi}(Q^{2} )& = m_\pi \sum_{j=1}^{2}\left[E_{\pi}^{2} T_{j}^{\pi E E}(Q^{2})+E_{\pi} F_{\pi} T_{j}^{\pi E F}(Q^{2})\right. \\
& \qquad \left.+F_{\pi}^{2} T_{j}^{\pi F F}(Q^{2})\right] \mathcal{V}_{j}(Q^{2}),
\label{AlgebraFpiT}
\end{aligned}
\end{equation}
where $\mathcal{V}_{i}(Q^{2};\zeta)$ are given in Eqs.\,(\ref{V1}) and (\ref{V2}), and the pion Bethe-Salpeter amplitude factors are given in Eq.\,\eqref{pionBSA}.  The coefficient functions obtained using the distinct regularisation schemes are recorded below.

\subsection{Scheme I}
{\allowdisplaybreaks
\begin{align}
    T^{\pi XY}_{I j}(Q^{2})& =\frac{3}{8\pi^{2}}
    \bigg\{\int_{0}^{1} d\alpha \, \mathcal{N}_{j}^{\pi XY}\bar{\mathcal{C}}^{\mathrm{iu}}_{1}(\omega) \nonumber \\
    & \quad + \int_{0}^{1} d\alpha d\beta \  2\alpha\bigg[\mathcal{A}_{j}^{\pi XY} \bar{\mathcal{C}}^{\mathrm{iu}}_{1}(\omega_{\pi}) \nonumber \\
    & \qquad + (\mathcal{B}_{j}^{\pi XY}-\omega_{\pi}\mathcal{A}_{j}^{\pi  X Y})\frac{\bar{\mathcal{C}}^{\mathrm{iu}}_{2}(\omega_{\pi})}{\omega_{\pi}}\bigg] \bigg\}\,,
\label{Tpion2}
\end{align}
with $j=1,2$, $\{X,Y\}=\{E, F\}$,
\begin{subequations}
\begin{align}
    \omega &=M^{2}+\alpha(1-\alpha)Q^{2},\\
    \omega_{\pi} &=M^{2}+Q^{2}\alpha^{2}\beta(1-\beta)-\alpha(1-\alpha)m_{\pi}^{2}\,,
    \label{omegapi2}
\end{align}
\end{subequations}
and, for $j=1$,
\begin{subequations}
\begin{align}
    \mathcal{A}_{1}^{\pi  EE}&=0 = \mathcal{N}_{1}^{\pi  EE}\,, \\
    \mathcal{B}_{1}^{\pi  EE}&=4 M \,,\\
    \mathcal{N}_{1}^{\pi  EF}&=8/M\,, \\
    \mathcal{A}_{1}^{\pi EF}& = -\frac{2  (3\alpha+2)}{M}\,\\
    \mathcal{B}_{1}^{\pi  EF}&=-\frac{4 }{M} \left\{M^2 (\alpha+4)\right.\nonumber\\
    & \qquad +\alpha \left[Q^{2} \alpha (\alpha+2) (\beta-1) \beta \right. \nonumber \\
    & \qquad \quad \left.\left. - m_{\pi}^2 \left(\alpha^2+2 \alpha-1\right)\right]\right\}\,,\\
    \mathcal{N}_{1}^{\pi  FF} &= -12/M\,,\\
    \mathcal{A}_{1}^{\pi  FF}&=8/M\,\\
    \mathcal{B}_{1}^{\pi, FF}&=16 M  \nonumber \\
    & \quad +\frac{8 \alpha \left(m_{\pi}^2 (1-2 \alpha)+2 Q^{2} \alpha (\beta-1) \beta\right)}{M}, \\
    \end{align}
\end{subequations}
 whereas with $j=2$,
\begin{subequations}
\begin{align}
    \mathcal{N}_{2}^{\pi  EE}&=- 16 /M\,,\\
    \mathcal{A}_{2}^{\pi  EE}&=12 \alpha/M\,,\\
    \mathcal{B}_{2}^{\pi  EE}&=\frac{8  \alpha \left(M^2-m_{\pi}^2 \left(\alpha^2+1\right)+Q^{2} \alpha^2 (\beta-1) \beta\right)}{M}\,,\\
    \mathcal{N}_{2}^{\pi EF}&= 16/M\,,\\
    \mathcal{A}_{2}^{\pi EF}&=0\,,\\
    \mathcal{B}_{2}^{\pi EF}&=\frac{8 (2 m_{\pi}^2-Q^{2} \alpha)}{M}\,,\\
    \mathcal{N}_{2}^{\pi  FF}&=\frac{16 m_{\pi}^2}{M^3}\,,\\
    \mathcal{A}_{2}^{\pi  FF}&=-\frac{4 \left(m_{\pi}^2 (3 \alpha+2)+Q^{2} \alpha\right)}{M^3}\,,\\
    \mathcal{B}_{2}^{\pi  FF} &= \frac{8  \alpha}{M^3} \bigg\{ M^2 \left(m_{\pi}^2+Q^{2}\right) \nonumber\\
     & \qquad -m_{\pi}^2 \left[ m_{\pi}^2 \left(\alpha^2-4 \alpha+1\right) \right.\nonumber \\
     & \qquad\quad \left.+ Q^{2} \alpha \left(3 \alpha \beta^2-3 \alpha \beta+\alpha-1\right) \right] \bigg\}\,\\
\end{align}
\end{subequations}

\subsection{Scheme II}
\allowdisplaybreaks
\begin{align}
    & T^{\pi XY}_{II j}(Q^{2})
    = \frac{3}{8\pi^{2}}\int d\alpha d\beta \,  2\alpha
    \bigg[
    (\mathcal{A}_{j}^{\pi  X Y} \! + \tfrac{1}{4}\mathcal{N}_{j}^{\pi X Y}) \mathsf{C}^{\mathrm{iu}}_{0}(\omega_{\pi}) \nonumber \\
    &   \qquad \quad   + (\mathcal{B}_{j}^{\pi, X Y}
    -\omega_{\pi}\mathcal{A}_{j}^{\pi X Y}) \mathsf{C}^{\mathrm{iu}}_{-2}(\omega_{\pi})\bigg]
\label{Tpion}
\end{align}
with $\omega_{\pi}$ given in Eq.\,\eqref{omegapi2} and, for $j=1$,
\begin{subequations}
\begin{align}
    \mathcal{A}_{1}^{\pi  EE} &=0 = \mathcal{N}_{1}^{\pi, EE} , \\
    \mathcal{B}_{1}^{\pi EE}&= 4 M \,,\\
    \mathcal{A}_{1}^{\pi EF}&=\frac{4(2-\alpha)}{M}\,,\\
    \mathcal{N}_{1}^{\pi EF}&=-\frac{8  (\alpha+2)}{M}\,,\\
    \mathcal{B}_{1}^{\pi EF}&=-\frac{4 }{M} \bigg[M^2 (\alpha+2) \nonumber \\
    & \qquad +m_{\pi}^2 \alpha (1-\alpha^2)+Q^{2} \alpha^3 (\beta-1) \beta\bigg]\,,\\
    \mathcal{A}_{1}^{\pi FF}&=-\frac{12}{M}\,,\\
    \mathcal{N}_{1}^{\pi FF}&=\frac{32}{M}\,,\\
    \mathcal{B}_{1}^{\pi FF}&=\frac{4 }{M} \left[M^2+\alpha (Q^{2} \alpha (\beta-1) \beta-m_{\pi}^2 (\alpha-2))\right]\,,
    \end{align}
\end{subequations}
whereas with $j=2$,
\begin{subequations}
\begin{align}
    \mathcal{A}_{2}^{\pi EE}&=\frac{8  (\alpha-2)}{M}\,,\\
    \mathcal{N}_{2}^{\pi  EE}&=\frac{16 \alpha}{M}\,,\\
    \mathcal{B}_{2}^{\pi EE}&=\frac{8 }{M} \bigg[M^2 (\alpha-2)+Q^{2} \alpha^{2} (\alpha-2) (\beta-1) \beta \nonumber \\
    & \qquad \quad -m_{\pi}^2 \alpha (\alpha-1)^2\bigg]\,,\\
    \mathcal{A}_{2}^{\pi EF}&=\frac{16 }{M},\\\quad
    \mathcal{N}_{2}^{\pi  EF}&=0\,,\\
    \mathcal{B}_{2}^{\pi  EF}&=\frac{8 }{M}\bigg[2 M^2+2 m_{\pi}^2 (1-\alpha^2) \nonumber \\
    & \qquad \quad +Q^{2} \alpha (2 \alpha (\beta-1) \beta-1)\bigg]\,,\\
    \mathcal{A}_{2}^{\pi FF}&=-\frac{8 }{M^3}\left[m_{\pi}^2 (3 \alpha-2)+Q^{2} \alpha\right]\,,\\
    \mathcal{N}_{2}^{\pi  FF}&=\frac{16 }{M^3}\left[m_{\pi}^2 (3 \alpha-2)+Q^{2} \alpha\right]\,,\\
    \mathcal{B}_{2}^{\pi FF}&=\frac{8 }{M^{3}}
    \bigg[M^2 (m_{\pi}^2 (\alpha-2)+Q^{2} \alpha)
    \nonumber \\
    & \quad -m_{\pi}^2 \alpha
        \bigg(m_{\pi}^2 (\alpha-1)^2
         +Q^{2} \alpha \big[ 3 \alpha \beta^2 \nonumber\\
    & \qquad  -3 \alpha \beta+\alpha-2 \beta^2+2 \beta-1\big] \bigg)\bigg]\,.
\end{align}
\end{subequations}
}

\section{Rho Tensor Form Factor}
\label{ApprhoFT}
Each $u$-in-$\rho$ tensor form factor can be written in the form:
\begin{align}
    F_{T l}^{\rho}\left(Q^{2}\right) & = E_{\rho}^{2} \sum^{2}_{j=1} T_{j}^{\rho l}\left(Q^{2}\right) \mathcal{V}_{j}\left(Q^{2}\right)\,,
\end{align}
where
the coefficient functions obtained using the distinct regularisation schemes are recorded below,
with
\begin{equation}
\omega_{\rho} =M^{2}+Q^{2}\alpha^{2}\beta(1-\beta)-\alpha(1-\alpha)m_{\rho}^{2}\,.
\end{equation}

\subsection{Scheme I}
{\allowdisplaybreaks
\begin{align}
        T_{I j l}^{\rho}(Q^{2})&=\frac{3}{8\pi^{2}}\bigg\{\int_{0}^{1} d\alpha \ \mathcal{N}_{jl}^{\rho}\bar{\mathcal{C}}^{\mathrm{iu}}_{1}(\omega) \nonumber \\
    & \quad + \int_{0}^{1} d\alpha d\beta\ 2\alpha
    \bigg[ \mathcal{A}_{jl}^{\rho}\bar{\mathcal{C}}^{\mathrm{iu}}_{1}(\omega_{\rho})\nonumber\\
    & \qquad + (\mathcal{B}_{jl}^{\rho} -\omega_{\rho} \mathcal{A}_{jl}^{\rho}) \frac{\bar{\mathcal{C}}^{\mathrm{iu}}_{2}(\omega_{\rho})}{\omega_{\rho}}\bigg]\bigg\}\,.
    \label{rhoF1}
\end{align}
Coefficient functions.\\
$l_I=1$.
\begin{subequations}
\label{rhoF1A}
\label{rhoj1*}
    \begin{align}
        \mathcal{N}_{1 1}^{\rho}&=\frac{4  M}{m_{\rho}}\,,\\
        \mathcal{A}_{11}^{\rho}&=0 = \mathcal{A}_{21}^{\rho} = \mathcal{B}_{2 1}^{\rho} = \mathcal{N}_{2 1}^{\rho}\,,\\
        \mathcal{B}_{1 1}^{\rho}&=\frac{2M(2m_{\rho}^{2}+Q^{2})}{m_{\rho}}\,.
    \end{align}
\end{subequations}
$l_I=2$.
\begin{subequations}
\label{rhoj2*}
    \begin{align}
        \mathcal{N}_{1 2}^{\rho}&=\frac{16  M}{m_{\rho}}\,,\\
        \mathcal{A}_{1 2}^{\rho}&=0 = \mathcal{A}_{2 2}^{\rho} = \mathcal{B}_{2 2}^{\rho} = \mathcal{N}_{2 2}^{\rho} \\
        \mathcal{B}_{1 2}^{\rho}&=\frac{8M(2m_{\rho}^{2}+Q^{2})\alpha}{m_{\rho}}\,.
    \end{align}
\end{subequations}
$l_I=3$.
\begin{subequations}
\label{rhoj3*}
\begin{align}
        \mathcal{N}_{1 3}^{\rho}&=\frac{4  M}{m_{\rho}}\,,\\
        \mathcal{A}_{1 3}^{\rho}&=0\,, \\
        \mathcal{B}_{1 3}^{\rho}&=-\frac{2M(2m_{\rho}^{2}+Q^{2})\alpha}{m_{\rho}}\,,\\
        \mathcal{N}_{2 3}^{\rho}&=\frac{8(2m_{\rho}^{2}+Q^{2})}{M m_{\rho}}\,,\\
        \mathcal{A}_{2 3}^{\rho}&=\frac{2\left(2m_{\rho}^{2}+Q^{2}\right)(\alpha-2)}{M m_{\rho}}\,\\
        \mathcal{B}_{2 3}^{\rho} & =\frac{1}{M m_\rho}
        \left[4 \alpha \left(2 m_{\rho}^2+Q^{2}\right) \right. \nonumber \\
        & \quad \times \left[M^2+m_{\rho}^2 \left(-\alpha^2 +2 \alpha+1\right) \right.\nonumber \\
        & \left. \qquad  +Q^{2} (\alpha-2) \alpha (\beta-1) \beta\right]\,.
\end{align}
\end{subequations}
$l_I=4$.
\begin{subequations}
\label{rhoj4*}
    \begin{align}
        \mathcal{N}_{1 4}^{\rho}&=0 = \mathcal{A}_{1 4}^{\rho} = \mathcal{N}_{2,4}^{\rho} \,,\\
        \mathcal{B}_{1 4}^{\rho}&=-4 M m_{\rho}\,,\\
        \mathcal{A}_{2 4}^{\rho}&=-\frac{4 m_{\rho} (\alpha-2)}{M}\,,\\
        \mathcal{B}_{2 4}^{\rho}&= \frac{8 m_{\rho}}{M}
        \left[ \alpha \left(m_{\rho}^2 (\alpha-1)^2 \right. \right.\nonumber \\
        & \left. \left. \quad - Q^{2} (\alpha-2) \alpha (\beta-1) \beta\right)-M^2 (\alpha-2)\right]\,.
    \end{align}
\end{subequations}
$l_I=5$.
\begin{subequations}
\label{rhoj5*}
    \begin{align}
        \mathcal{N}_{1 5}^{\rho}&= \frac{16 M}{m_{\rho}}\,,\\
        \mathcal{A}_{1 5}^{\rho}&=0 =\mathcal{B}_{1 5}^{\rho} = \mathcal{N}_{2,5}^{\rho}\\
        \mathcal{A}_{2 5}^{\rho}&=\frac{16 \alpha \left(2 m_{\rho}^2+Q^{2}\right)}{M m_{\rho}}\,,\\
        \mathcal{B}_{2 5}^{\rho}&=
        \frac{32 \alpha  \left(2 m_{\rho}^2+Q^{2}\right)}{M m_{\rho}} \nonumber \\
        & \quad \times
        \left[M^2-m_{\rho}^2 (\alpha-1) \alpha \left(2 \beta^2-2 \beta+1\right)\right]\,.
    \end{align}
\end{subequations}
}

\subsection{Scheme II}
{\allowdisplaybreaks
\begin{align}
        T_{II j l}^{\rho}(Q^{2}) & = \frac{3}{8\pi^{2}} \int d\alpha d\beta \,
        2\alpha\left[ (\mathcal{A}_{jl}^{\rho} +\mathcal{N}_{j l }^{\rho}/4) \mathsf{C}^{\mathrm{iu}}_{0}(\omega_{\rho})\right. \nonumber \\
        &\qquad \left. +(\mathcal{B}_{j l }^{\rho} - \omega_{\rho} \mathcal{A}_{jl}^{\rho}) \mathsf{C}^{\mathrm{iu}}_{-2}(\omega_{\rho}) \right]\,.
        \label{rhoF1N}
\end{align}
Coefficient functions.\\
$l_{II}=1$.
\begin{subequations}
\label{rhoj1}
    \begin{align}
        \mathcal{N}_{1 1}^{\rho}& = 0 =
        \mathcal{A}_{2 1}^{\rho} =
        \mathcal{B}_{2 1}^{\rho} =
        \mathcal{N}_{2 1}^{\rho}\,,\\
        \mathcal{A}_{1 1}^{\rho}&=\frac{4  M}{m_{\rho}}\,,\label{B:rho11} \\
        \mathcal{B}_{1 1}^{\rho}&=\frac{2M}{m_{\rho}}\left[ 2 M^2-2 m_{\rho}^2 \left(\alpha^2-1\right)
        \nonumber \right.\\
        & \left. \quad +2 Q^{2} \alpha^2 (\beta-1) \beta+Q^{2}\right]\,.
\end{align}
\end{subequations}
$l_{II}=2$.
\begin{subequations}
\label{rhoj2}
\begin{align}
        \mathcal{N}_{1,2}^{\rho}&=0=
        \mathcal{A}_{2,2}^{\rho} =
        \mathcal{B}_{2,2}^{\rho} =
        \mathcal{N}_{2,2}^{\rho}\,,\\
        \mathcal{A}_{1,2}^{\rho}&=\frac{16  M}{m_{\rho}}\,,\\
        \mathcal{B}_{1,2}^{\rho}&=\frac{8M}{m_{\rho}}\left[2 M^2+\alpha \left(-2 m_{\rho}^2 (\alpha-1)
        \right.\right. \nonumber \\
        & \left.\left. \quad +2 Q^{2} \alpha (\beta-1) \beta+Q^{2}\right)\right]\,.
\end{align}
\end{subequations}
$l_{II}=3$.
\begin{subequations}
\label{rhoj3}
\begin{align}
        \mathcal{N}_{1 3}^{\rho}&=0\,,\\
        \mathcal{A}_{1 3}^{\rho}&=\frac{4  M}{m_{\rho}}\,,\label{B:rho13} \\
        \mathcal{B}_{1 3}^{\rho}&=\frac{2M}{m_{\rho}}
        [2 M^2-\alpha  ( 2 m_{\rho}^2 (\alpha+1) \nonumber \\
        & \quad -2 Q^{2} \alpha (\beta-1) \beta+Q^{2} ) ]\,, \\
        \mathcal{N}_{2 3}^{\rho}&=-\frac{8(\alpha+2) \left(2 m_{\rho}^2+Q^{2}\right)}{M m_{\rho}}\,,\\
        \mathcal{A}_{2 3}^{\rho}&=\frac{4 (\alpha+2) \left(2 m_{\rho}^2+Q^{2}\right)}{M m_{\rho}}\,,\\
        \mathcal{B}_{2 3}^{\rho}&=\frac{4\left(2 m_{\rho}^2+Q^{2}\right)}{M m_{\rho}}\
        [M^2 (\alpha+2)+m_{\rho}^2 \alpha (1-\alpha^2) \nonumber \\
        & \quad +Q^{2} \alpha^3 (\beta-1) \beta ]\,.
\end{align}
\end{subequations}
$l_{II}=4$.
\begin{subequations}
\label{rhoj4}
    \begin{align}
        \mathcal{N}_{1 4}^{\rho}&=0 = \mathcal{A}_{1 4}^{\rho}\,,\\
        \mathcal{B}_{1 4}^{\rho}&=-4 M m_{\rho},\\\quad
        \mathcal{N}_{2,4}^{\rho}&=\frac{16 m_{\rho} (\alpha-2)}{M} \,,\\
        \mathcal{A}_{2 4}^{\rho}&=\frac{8 m_{\rho} (2-\alpha)}{M}\,,\\
        \mathcal{B}_{2 4}^{\rho}&=\frac{8m_{\rho}}{M} [\alpha (m_{\rho}^2 (\alpha-1)^2-Q^{2} (\alpha-2) \alpha (\beta-1) \beta) \nonumber \\
        & \quad -M^2 (\alpha-2)]\,.
    \end{align}
\end{subequations}
$l_{II}=5$.
\begin{subequations}
\label{rhoj5}
    \begin{align}
        \mathcal{N}_{1 5}^{\rho}&=0\,,\\
        \mathcal{A}_{1 5}^{\rho}&=\frac{16 M}{m_{\rho}}\,\\
        \mathcal{B}_{1 5}^{\rho}&=\frac{16 M}{m_{\rho}} \nonumber \\
        & \quad \times \left[M^2-\alpha^2 \left(m_{\rho}^2-Q^{2} (\beta-1) \beta\right)\right]\,,\\
        \mathcal{N}_{2 5}^{\rho}&=-\frac{64\alpha(2 m_{\rho}^{2}+Q^{2})}{M m_{\rho}}\,,\\
        \mathcal{A}_{2 5}^{\rho}&=\frac{32\alpha(2 m_{\rho}^{2}+Q^{2})}{M m_{\rho}}\,,\\
        \mathcal{B}_{2 5}^{\rho}&=\frac{32  \alpha \left(2 m_{\rho}^{2}+Q^{2}\right)}{M m_{\rho}} [M^{2}  \nonumber \\
        & \quad -m_{\rho}^{2} (\alpha-1) \alpha (2 \beta^{2}-2 \beta+1)]\,. \label{B:rho25}
    \end{align}
\end{subequations}
}

\subsection{Rho tensor form factor interpolations}
\label{InterprhoT}
With $z=Q^2$, these interpolations are valid at the resolving scale $\zeta=\zeta_2$.
{\allowdisplaybreaks
\begin{subequations}
    \begin{align}
        F^{\rho {\rm I}}_{T 1}(z)& =
        \frac{0.499+0.268 z +0.020z^2}{1+0.547 z+0.032 z^2} = F^{\rho {\rm II}}_{T 1}(z)\\
        %
        F^{\rho {\rm I}}_{T 2}(z) & =
        \frac{1.433+0.531 z +0.024 z^2}{1+0.542 z + 0.028z^2 }= F^{\rho {\rm II}}_{T 2}(z) \label{rhoT2*}\,,\\
        F^{\rho{\rm I}}_{T 3}(z)&=
        \frac{-1.078-0.659 z -0.050 z^2}{1+1.795 z +0.229 z^2}
        \label{rhoT3*}\,,\\
        F^{\rho{\rm I}}_{T 4}(z)&=
        \frac{-0.839-0.298 z - 0.002 z^2}{1+1.917 z + 0.473 z^2 }
        \label{rhoT4*}\,,\\
        F^{\rho{\rm I}}_{T 5}(z)& =
        \frac{-1.343 + 0.002 z + 0.004z^2}{1+1.745 z - 0.081 z^2}
        \,.\label{rhoT5*}
    \end{align}
\end{subequations}
\begin{subequations}
    \begin{align}
        F^{\rho{\rm II}}_{T 3}(z)&=
        \frac{-1.048-0.628 z-0.075 z^2}{1+1.862 z + 0.340 z^2}
        \label{rhoT3}\,,\\
        F^{\rho{\rm II}}_{T 4}(z)&=
       \frac{-0.827 - 0.154 z - 0.001 z^2}{1+1.790 z + 0.231 z^2}
        \label{rhoT4}\,,\\
        F^{\rho{\rm II}}_{T 5}(z)& =
        \frac{-1.271-0.316 z + 0.001 z^2}{1+2.133 z + 0.760 z^2}
        \,.\label{rhoT5}
    \end{align}
\end{subequations}
}

\section{Pi-rho transition tensor form factors}
\label{Apppirho}
Each $u$-quark $\pi\leftrightarrow\rho$ tensor transition form factor can be written in the form:
\begin{align}
    F_{T l}^{\pi\rho}\left(Q^{2}\right) & = \sum^{2}_{j=1} T_{jl}^{\pi \rho }\left(Q^{2}\right) \mathcal{V}_{j}\left(Q^{2}\right)\,,
\end{align}
where the quark tensor vertex form factors are discussed in Sect.\,\ref{QTV},
and
the coefficient functions obtained using the distinct regularisation schemes are recorded below, with
\begin{align}
\omega_{\pi\rho} & =M^2+\alpha \left[m_{\pi}^2 (\alpha-1) \beta - m_{\rho}^2 (\alpha\beta-\alpha-\beta+1)\right. \nonumber \\
& \left. \quad -Q^{2} \alpha \beta (\beta-1)\right]\,.
\end{align}

\subsection{Scheme I}
{\allowdisplaybreaks
\begin{align}
    T^{\pi\rho }_{I j  l}&(Q^{2})=\frac{3}{8\pi^{2}}\int_{0}^{1} d\alpha d\beta \  2\alpha\bigg[\mathcal{A}_{i,j}^{\pi\rho}\bar{\mathcal{C}}^{\mathrm{iu}}_{1}(\omega_{\pi\rho})
    \nonumber\\
    & \qquad \quad +(\mathcal{B}_{i,j}^{\pi\rho}- \omega_{\pi\rho} \mathcal{A}_{i,j}^{\pi\rho}) \bar{\mathcal{C}}^{\mathrm{iu}}_{2}(\omega_{\pi\rho})/\omega_{\pi\rho}\bigg]\,.
    \label{Tpionrho2*}
\end{align}
Coefficient functions.\\
$l_I=1$.
\begin{subequations}
    \begin{align}
    \mathcal{A}_{1 1}^{\pi\rho}&=6 E_{\pi} E_{\rho}(\alpha-1)\,,\\
    \mathcal{B}_{1 1}^{\pi\rho}&=4 E_{\rho}[
    E_{\pi} M^2 (\alpha-2) \nonumber \\
    &  \qquad + E_{\pi} \alpha \{-m_{\pi}^2 (\alpha-1)^2 \beta \nonumber\\
    & \qquad \quad + m_{\rho}^2 (\alpha-1)^2 (\beta-1)+Q^{2} (\alpha-2) \alpha (\beta-1) \beta\}\nonumber\\
    &\qquad+F_{\pi} \{M^2-m_{\pi}^2 (\alpha-1) (\alpha \beta+1)+\alpha (\beta-1)\nonumber\\
    & \qquad \quad \times [m_{\rho}^2 (\alpha-1)+Q^{2} \alpha \beta+Q^{2}]\}]\,,\\
    \mathcal{A}_{2 1}^{\pi\rho}&=-\frac{4 E_{\rho}F_{\pi} Q^{2} (3 \alpha (\beta-1)+1)}{M^2}\,,\\
    \mathcal{B}_{2 1}^{\pi\rho}&=- \frac{8 E_{\rho}Q^{2}}{M^2} [E_{\pi} M^2+F_{\pi} M^2 (\alpha (\beta-1)-1)\nonumber\\
    & \qquad +F_{\pi} \alpha \{m_{\pi}^2 \beta (\alpha^2 (1-\beta)+\alpha \beta-1)\nonumber\\
    & \qquad \quad +m_{\rho}^2 (\alpha-1) \alpha (\beta-1)^2\nonumber \\
    & \qquad \qquad +Q^{2} \alpha^2 (\beta-1)^2 \beta\}]\,.
    \end{align}
\end{subequations}
$l_I=2$.
\begin{subequations}
    \begin{align}
    \mathcal{A}_{1 2}^{\pi\rho} & =E_{\pi} E_{\rho}(\alpha (6 \beta-3)+1)\,,\\
    \mathcal{B}_{1 2}^{\pi\rho} & = 2 E_{\rho}[
    E_{\pi} \alpha \{M^2 (2 \beta-1) \nonumber\\
    & \qquad \quad +\beta (m_{\pi}^2 (-2 \alpha^2 \beta+\alpha^2+1)\nonumber\\
    & \qquad \qquad +Q^{2} \alpha^2 (2 \beta^2-3 \beta+1)) \nonumber \\
    & \qquad \quad +m_{\rho}^2 (\beta-1) (\alpha^2 (2 \beta-1)+1)\}\nonumber\\
    &\qquad + F_{\pi} \{M^2-m_{\pi}^2 (\alpha^2 \beta+\alpha (\beta-1)+1) \nonumber \\
    & \qquad \quad +\alpha (\beta-1) [m_{\rho}^2 (\alpha-3) \nonumber \\
    & \qquad \qquad +Q^{2} (\alpha \beta-1)]\}],\\\quad
    \mathcal{A}_{2 2}^{\pi\rho}&= \frac{2 E_{\rho}F_{\pi} (m_{\rho}^2-m_{\pi}^2) (3 \alpha (\beta-1)+1)}{M^2}\,,\\
    \mathcal{B}_{2 2}^{\pi\rho}&=
    \frac{4 E_{\rho}}{M^2}(m_{\rho}^2-m_{\pi}^2) [E_{\pi} M^2 \nonumber \\
    & \quad +F_{\pi} M^2 (\alpha (\beta-1)-1)\nonumber\\
    &\quad -F_{\pi} \alpha \{m_{\pi}^2 \beta (\alpha^2 (\beta-1) - \alpha \beta+1)\nonumber\\
    &\qquad +m_{\rho}^2 (\alpha-1) \alpha (\beta-1)^2 \nonumber \\
    & \qquad \quad +Q^{2} \alpha^2 (\beta-1)^2 \beta\}]\,.
    \end{align}
\end{subequations}
$l_I=3$.
\begin{subequations}
    \begin{align}
    \mathcal{A}_{1 3}^{\pi\rho}&=0\,,\\
    \mathcal{B}_{1 3}^{\pi\rho}&=8 E_{\rho}m_{\rho}^2 \alpha^2 (1-\beta) \beta (E_{\pi}-2 F_{\pi})\,,\\
    \mathcal{A}_{2 3}^{\pi\rho}&=-\frac{4 E_{\rho}F_{\pi} m_{\rho}^2 (3 \alpha (\beta-1)+1)}{M^2}\,,\\
    \mathcal{B}_{2 3}^{\pi\rho}&=-\frac{8 E_{\rho}m_{\rho}^2}{M^2}
    [E_{\pi} M^2 \nonumber \\
    & \qquad +F_{\pi} M^2 (\alpha (\beta-1)-1)\nonumber\\
    &\qquad +F_{\pi} \alpha \{m_{\pi}^2 \beta (\alpha \beta -\alpha^2 (\beta-1)-1) \nonumber \\ %
    & \qquad \quad +m_{\rho}^2 (\alpha-1) \alpha (\beta-1)^2\nonumber\\
    & \qquad \quad +Q^{2} \alpha^2 (\beta-1)^2 \beta\}]\,.
    \end{align}
\end{subequations}
}

\subsection{Scheme II}
{\allowdisplaybreaks
\begin{align}
        T_{II j l }^{\pi\rho}&(Q^{2})=\frac{3}{8\pi^{2}}\int d\alpha d\beta  2\alpha [(\mathcal{A}_{j l }^{\pi\rho}+\mathcal{N}_{j l }^{\pi\rho}/4) \mathsf{C}^{\mathrm{iu}}_{0}(\omega_{\pi\rho})\nonumber \\
    & \qquad \quad +(\mathcal{B}_{j l }^{\pi\rho}-\omega_{\pi\rho}\mathcal{A}_{j l }^{\pi\rho}) \mathsf{C}^{\mathrm{iu}}_{-2}(\omega_{\pi\rho})]\,.
\label{Tpionrho2}
\end{align}
%
Coefficient functions.\\
$l_{II}=1$.
\begin{subequations}
    \begin{align}
    \mathcal{A}_{1 1}^{\pi\rho} & = 4 E_{\pi} E_{\rho}(\alpha-1) \,, \\
    \mathcal{N}_{1 1}^{\pi\rho} & = 2 \mathcal{A}_{1 1}^{\pi\rho}\,,\\
    \mathcal{B}_{1 1}^{\pi\rho} & =4 E_{\rho} [E_{\pi} M^2 (\alpha-2) \nonumber \\
    & \quad +E_{\pi} \alpha \{ - m_{\pi}^2 (\alpha-1)^2 \beta
    +m_{\rho}^2 (\alpha-1)^2 (\beta-1)\nonumber\\
    & \qquad + Q^{2} (\alpha-2) \alpha (\beta-1) \beta\} \nonumber \\
    & \quad +F_{\pi}\{M^2-m_{\pi}^2 (\alpha-1) (\alpha \beta+1)+\alpha (\beta-1)\nonumber\\
    &\qquad \quad \times(m_{\rho}^2 (\alpha-1)+Q^{2} \alpha \beta+Q^{2})\}]\,,\\
    \mathcal{A}_{2 1}^{\pi\rho} & =\frac{E_{\rho} F_{\pi}}{M^2}
     [3 (m_{\rho}^2-m_\pi^2) (\alpha-1)\nonumber\\
    & \qquad +Q^{2} (-2 \alpha \beta+5 \alpha-3)]\,,\\
    \mathcal{N}_{2 1}^{\pi\rho}&=- \frac{4 E_{\rho} F_{\pi}}{M^2}
     [3 (m_{\rho}^2 -m_\pi^2) (\alpha-1)\nonumber\\
    &\qquad +Q^{2} \alpha (10 \beta-7)+Q^{2}]\,,\\
    \mathcal{B}_{2 1}^{\pi\rho}&= - \frac{8 E_{\rho} Q^{2}}{M^2} [E_{\pi} M^2 \nonumber \\
    & \quad +F_{\pi} M^2 (\alpha (\beta-1)-1)\nonumber\\
    & \quad +F_{\pi} \alpha \{m_{\pi}^2 \beta (-(\alpha^2 (\beta-1))+\alpha \beta-1) \nonumber \\
    & \qquad +m_{\rho}^2 (\alpha-1) \alpha (\beta-1)^2\nonumber\\
    &\qquad \quad +Q^{2} \alpha^2 (\beta-1)^2 \beta\}]\,.
    \end{align}
\end{subequations}
$l_{II}=2$.
\begin{subequations}
    \begin{align}
    \mathcal{A}_{1 2}^{\pi\rho}&=2 E_{\pi} E_{\rho} (\alpha (2 \beta-1)+1)\,,\\
    \mathcal{N}_{1 2}^{\pi\rho}&= 4 E_{\pi} E_{\rho} (\alpha (2 \beta-1)-1)\,,\\
    \mathcal{B}_{1 2}^{\pi\rho}&=2 E_{\rho} [E_{\pi} \alpha [M^2 (2 \beta-1) \nonumber \\
    & \quad + \beta \{m_{\pi}^2 (-2 \alpha^2 \beta+\alpha^2+1)\nonumber\\
    & \qquad +Q^{2} \alpha^2 (2 \beta^2-3 \beta+1)\} \nonumber \\
    & \quad +m_{\rho}^2 (\beta-1) (\alpha^2 (2 \beta-1)+1)]\nonumber\\
    &\quad +F_{\pi} \{M^2-m_{\pi}^2 (\alpha^2 \beta+\alpha (\beta-1)+1)\nonumber \\
    & \qquad +\alpha (\beta-1) (m_{\rho}^2 (\alpha-3)+Q^{2} (\alpha \beta-1))\}]\,,\\
    \mathcal{A}_{2 2}^{\pi\rho}&= \frac{E_{\rho} F_{\pi}}{2 M^2}[m_{\pi}^2 (\alpha (7-6 \beta)-5)+m_{\rho}^2 (\alpha (6 \beta-3)+1)\nonumber\\
    & \qquad +Q^{2} \alpha (4 \beta-1)+Q^{2}]\,,\\
    \mathcal{N}_{2 2}^{\pi\rho}&=-\frac{2 E_{\rho} F_{\pi}}{M^2}
        [m_{\pi}^2 (\alpha (6 \beta-5)-1) \nonumber \\
        & \qquad +m_{\rho}^2 (\alpha (9-6 \beta)-3)\nonumber\\
       & \qquad \quad + Q^{2} (\alpha (4 \beta-1)+1)]\,,\\
    \mathcal{B}_{2 2}^{\pi\rho}&=-\frac{4 E_{\rho}}{M^2} (m_{\pi}^2-m_{\rho}^2)
    [E_{\pi} M^2+F_{\pi} M^2 (\alpha (\beta-1)-1)\nonumber\\
    &\qquad +F_{\pi} \alpha \{m_{\pi}^2 \beta (-(\alpha^2 (\beta-1))+\alpha \beta-1) \nonumber \\
    & \qquad \quad +m_{\rho}^2 (\alpha-1) \alpha (\beta-1)^2\nonumber\\
    &\qquad \qquad +Q^{2} \alpha^2 (\beta-1)^2 \beta\}]\,.
    \end{align}
\end{subequations}
$l_{II}=3$.
\begin{subequations}
    \begin{align}
    \mathcal{A}_{1 3}^{\pi\rho}&=0 = \mathcal{N}_{1 3}^{\pi\rho}\,,\\
    \mathcal{B}_{1 3}^{\pi\rho}&=-8 E_{\rho} m_{\rho}^2 \alpha^2 (\beta-1) \beta [E_{\pi}-2 F_{\pi}]\,,\\
    \mathcal{A}_{2 3}^{\pi\rho}&=-\frac{2 E_{\rho} F_{\pi} m_{\rho}^2 (\alpha (3 \beta-4)+3)}{M^2},\\\quad
    \mathcal{N}_{2 3}^{\pi\rho}&=-\frac{8 E_{\rho} F_{\pi} m_{\rho}^2 (\alpha (3 \beta-2)-1)}{M^2}\,,\\
    \mathcal{B}_{2 3}^{\pi\rho}&=-\frac{8 E_{\rho} m_{\rho}^2}{M^2}
    [E_{\pi} M^2+F_{\pi} M^2 (\alpha (\beta-1)-1)\nonumber\\
    & \qquad +F_{\pi} \alpha \{m_{\pi}^2 \beta (\alpha \beta-\alpha^2 (\beta-1)-1)\nonumber \\
    & \qquad \quad +m_{\rho}^2 (\alpha-1) \alpha (\beta-1)^2\nonumber\\
    & \qquad\qquad +Q^{2} \alpha^2 (\beta-1)^2 \beta\}]\,.
    \end{align}
\end{subequations}
}

\subsection{Pi-rho transition tensor form factor interpolations}
\label{ApppirhoI}
With $z=Q^2$, the following interpolations are valid at the resolving scale $\zeta=\zeta_2$.
{\allowdisplaybreaks
\begin{subequations}
    \begin{align}
        F^{ \pi\rho I}_{T 1}(z)&=
        \frac{-0.859 - 0.081 z - 0.001 z^2}{1+1.373 z - 0.027 z^2}
        \label{rhopionT1*}\,,\\
        F^{\pi\rho I }_{T 2}(z)&=
        \frac{-0.155+0.156 z + 0.046 z^2}{1+1.937 z + 0.561 z^2}
        \label{rhopionT2*}\,,\\
        F^{\pi\rho I }_{T 3}(z)&=
        \frac{0.565+0.015 z - 0.001 z^2}{1+1.795 z +0.308 z^2}
        \label{rhopionT3*}\,.
    \end{align}
\end{subequations}
\begin{subequations}
    \begin{align}
        F^{\pi\rho II}_{T 1}(z)&=
        \frac{-0.888 - 0.112 z + 0.005 z^2}{1 + 1.353 z - 0.061 z^2}
        \label{rhopionT1},\\\quad
        F^{\pi\rho II}_{T 2}(z)&=
        \frac{-0.192 + 0.119 z - 0.005 z^2}{1+1.754 z - 0.083 z^2}
        \label{rhopionT2},\\\quad
        F^{\pi\rho II}_{T 3}(z)&=
        \frac{0.554 + 0.017 z - 0.001 z^2}{1+1.814 z + 0.328 z^2}
        \label{rhopionT3}\,.
    \end{align}
\end{subequations}
}

\section{Diquark tensor form factor interpolations}
\label{AppE}
As usual, $z=Q^2$, and the following interpolations are valid at the resolving scale $\zeta=\zeta_2$ on the domain $0\leq z \leq 10\,$GeV$^2$.

{\allowdisplaybreaks
\subsection{Scalar diquark elastic}
\begin{subequations}
\begin{align}
F^{0^{+} {\rm I}}_{T}(z)  & =
1.716 \frac{1+ 0.287 z + 0.00117 z^2}{1+1.810  z + 0.278 z^2 }\,,\\
F^{0^{+} {\rm I}}_{T}(z) & =
1.757\frac{1 + 0.111 z + 0.000285 z^2}{1 + 1.746 z + 0.149 z^2}\,.
\end{align}
\end{subequations}
Naturally, owing to Eq.\,\eqref{Tpinorm} and the larger mass of the scalar diquark compared to the $\pi$-meson, these form factors are roughly $5\approx m_{0^+}/m_\pi$-times greater in magnitude than $F_T^{\pi}$.

\subsection{Axialvector diquark elastic}
\begin{subequations}
    \begin{align}
        F^{1^+ {\rm I}}_{T 1}(z)& =
        \frac{0.435+0.169 z +0.014z^2}{1+0.520 z +0.039 z^2}= F^{1^+ {\rm II}}_{T 1}(z)\\
        %
        F^{1^+ {\rm I}}_{T 2}(z) & =
        \frac{1.133+1.295 z +0.244 z^2}{1+1.403 z + 0.451 z^2}= F^{1^+ {\rm II}}_{T 2}(z) \label{avT2*}\,,\\
        F^{1^+ {\rm I}}_{T 3}(z)&=
        \frac{-1.245-0.277 z -0.004z^2}{1+1.680 z + 0.036 z^2}
        \label{avT3*}\,,\\
        F^{1^+ {\rm I}}_{T 4}(z)&=
        \frac{-0.887-0.087 z -0.001 z^2}{1+1.702 z + 0.158 z^2}
        \label{avT4*}\,,\\
        F^{1^+ {\rm I}}_{T 5}(z)& =
        \frac{-1.775 - 1.078 z -0.005 z^2 }{1+2.430 z + 1.367 z^2}
         \,.\label{avT5*}
    \end{align}
\end{subequations}
\begin{subequations}
    \begin{align}
        F^{1^+ {\rm II}}_{T 3}(z)&=
        \frac{-1.288 - 0.827 z - 0.116 z^2}{1+2.167 z +0.856 z^2}
        \label{avT3}\,,\\
        \rule{0em}{5ex}
        F^{1^+ {\rm II}}_{T 4}(z)&=
        \frac{-0.915 - 0.504 z - 0.004 z^2}{1+2.217 z +0.963 z^2}
        \label{avT4}\,,\\
        F^{1^+ {\rm II}}_{T 5}(z)& =
        \frac{-1.837 - 0.438 z - 0.0003 z^2}{1+2.155 z + 0.828 z^2}
        \,.\label{avT5}
    \end{align}
\end{subequations}

\subsection{Scalar-axialvector diquark transition}
\begin{subequations}
    \begin{align}
        F^{ 0^+1^+ I}_{T 1}(z)&=
        \frac{-0.776 - 0.044 z -0.005 z^2}{1+1.370 z+0.028 z^2}
        \label{01T1*}\,,\\
        F^{0^+1^+ I }_{T 2}(z)&=
        \frac{0.016 + 0.161 z +0.032 z^2}{1 + 2.132 z +0.752 z^2}
        \label{01T2*}\,,\\
        \rule{0em}{5ex}
        F^{0^+1^+ I }_{T 3}(z)&=
        \frac{0.662 + 0.019 z -0.001 z^2}{1+1.858 z +0.452 z^2}
        \label{01T3*}\,.
    \end{align}
\end{subequations}
\begin{subequations}
    \begin{align}
        F^{0^+1^+ II}_{T 1}(z)&=
        \frac{-0.766 - 0.066 z -0.003 z^2}{1+1.387 z +0.014 z^2}
        \label{01T1}\,,\\
        F^{0^+1^+ II}_{T 2}(z)&=
        \frac{0.024 + 0.116 z +0.104 z^2}{1+1.639 z +2.856 z^2}
        \label{01T2}\,\\
        F^{0^+1^+ II}_{T 3}(z)&=
        \frac{0.664+0.021 z -0.001 z^2}{1+1.878 z +0.473 z^2}
        \label{01T3}\,.
    \end{align}
\end{subequations}
}


\begin{thebibliography}{89}
\expandafter\ifx\csname natexlab\endcsname\relax\def\natexlab#1{#1}\fi
\providecommand{\url}[1]{\texttt{#1}}
\providecommand{\href}[2]{#2}
\providecommand{\path}[1]{#1}
\providecommand{\DOIprefix}{doi:}
\providecommand{\ArXivprefix}{arXiv:}
\providecommand{\URLprefix}{URL: }
\providecommand{\Pubmedprefix}{pmid:}
\providecommand{\doi}[1]{\href{http://dx.doi.org/#1}{\path{#1}}}
\providecommand{\Pubmed}[1]{\href{pmid:#1}{\path{#1}}}
\providecommand{\bibinfo}[2]{#2}
\ifx\xfnm\relax \def\xfnm[#1]{\unskip,\space#1}\fi
\bibitem[{Fritzsch et~al.(1973)Fritzsch, Gell-Mann, and
  Leutwyler}]{Fritzsch:1973pi}
\bibinfo{author}{H.~Fritzsch}, \bibinfo{author}{M.~Gell-Mann},
  \bibinfo{author}{H.~Leutwyler},
\newblock \bibinfo{title}{{Advantages of the Color Octet Gluon Picture}},
\newblock \bibinfo{journal}{Phys. Lett. B} \bibinfo{volume}{47}
  (\bibinfo{year}{1973}) \bibinfo{pages}{365--368}.
\bibitem[{Pickering(1984)}]{Pickering:1984tk}
\bibinfo{author}{A.~Pickering}, \bibinfo{title}{{Constructing Quarks. A
  Sociological History of Particle Physics}}, \bibinfo{publisher}{University of
  Chicago Press}, \bibinfo{year}{1984}.
\bibitem[{Horn and Roberts(2016)}]{Horn:2016rip}
\bibinfo{author}{T.~Horn}, \bibinfo{author}{C.~D. Roberts},
\newblock \bibinfo{title}{{The pion: an enigma within the Standard Model}},
\newblock \bibinfo{journal}{J. Phys. G.} \bibinfo{volume}{43}
  (\bibinfo{year}{2016}) \bibinfo{pages}{073001}.
\bibitem[{Roberts et~al.(2021)Roberts, Richards, Horn, and
  Chang}]{Roberts:2021nhw}
\bibinfo{author}{C.~D. Roberts}, \bibinfo{author}{D.~G. Richards},
  \bibinfo{author}{T.~Horn}, \bibinfo{author}{L.~Chang},
\newblock \bibinfo{title}{{Insights into the emergence of mass from studies of
  pion and kaon structure}},
\newblock \bibinfo{journal}{Prog. Part. Nucl. Phys.} \bibinfo{volume}{120}
  (\bibinfo{year}{2021}) \bibinfo{pages}{103883}.
\bibitem[{Navas et~al.(2024)}]{ParticleDataGroup:2024cfk}
\bibinfo{author}{S.~Navas}, et~al.,
\newblock \bibinfo{title}{{Review of particle physics}},
\newblock \bibinfo{journal}{Phys. Rev. D} \bibinfo{volume}{110}
  (\bibinfo{year}{2024}) \bibinfo{pages}{030001}.
\bibitem[{Ding et~al.(2023)Ding, Roberts, and Schmidt}]{Ding:2022ows}
\bibinfo{author}{M.~Ding}, \bibinfo{author}{C.~D. Roberts},
  \bibinfo{author}{S.~M. Schmidt},
\newblock \bibinfo{title}{{Emergence of Hadron Mass and Structure}},
\newblock \bibinfo{journal}{Particles} \bibinfo{volume}{6}
  (\bibinfo{year}{2023}) \bibinfo{pages}{57--120}.
\bibitem[{de~Teramond(2022)}]{deTeramond:2022zcm}
\bibinfo{author}{G.~F. de~Teramond},
\newblock \bibinfo{title}{{Emergent phenomena in QCD: The holographic
  perspective -- arXiv:2212.14028 [hep-ph]}},
\newblock in: \bibinfo{booktitle}{{25th Workshop on What Comes Beyond the
  Standard Models?}}, \bibinfo{year}{2022}.
\bibitem[{Ferreira and Papavassiliou(2023)}]{Ferreira:2023fva}
\bibinfo{author}{M.~N. Ferreira}, \bibinfo{author}{J.~Papavassiliou},
\newblock \bibinfo{title}{{Gauge Sector Dynamics in QCD}},
\newblock \bibinfo{journal}{Particles} \bibinfo{volume}{6}
  (\bibinfo{year}{2023}) \bibinfo{pages}{312--363}.
\bibitem[{Raya et~al.(2024)Raya, Bashir, Binosi, Roberts, and
  Rodr\'\i{}guez-Quintero}]{Raya:2024ejx}
\bibinfo{author}{K.~Raya}, \bibinfo{author}{A.~Bashir},
  \bibinfo{author}{D.~Binosi}, \bibinfo{author}{C.~D. Roberts},
  \bibinfo{author}{J.~Rodr\'\i{}guez-Quintero},
\newblock \bibinfo{title}{{Pseudoscalar Mesons and Emergent Mass}},
\newblock \bibinfo{journal}{Few Body Syst.} \bibinfo{volume}{65}
  (\bibinfo{year}{2024}) \bibinfo{pages}{60}.
\bibitem[{Achenbach et~al.(2025)Achenbach, Carman, Gothe, Joo, Mokeev, and
  Roberts}]{Achenbach:2025kfx}
\bibinfo{author}{P.~Achenbach}, \bibinfo{author}{D.~S. Carman},
  \bibinfo{author}{R.~W. Gothe}, \bibinfo{author}{K.~Joo},
  \bibinfo{author}{V.~I. Mokeev}, \bibinfo{author}{C.~D. Roberts},
\newblock \bibinfo{title}{{Electroexcitation of Nucleon Resonances and the
  Emergence of Hadron Mass}},
\newblock \bibinfo{journal}{Symmetry} \bibinfo{volume}{17}
  (\bibinfo{year}{2025}) \bibinfo{pages}{1106}.
\bibitem[{Binosi(2026)}]{Binosi:2026tre}
\bibinfo{author}{D.~Binosi},
\newblock \bibinfo{title}{{Gauge Symmetry Beyond Perturbation Theory: BRST and
  anti-BRST Structure, Background Fields, and Infrared Dynamics of Yang--Mills
  Theory}},
\newblock \bibinfo{journal}{Particles} \bibinfo{volume}{9}
  (\bibinfo{year}{2026}) \bibinfo{pages}{59}.
\bibitem[{Cui et~al.(2020)Cui, Ding, Gao, Raya, Binosi, Chang, Roberts,
  Rodr\'{\i}guez-Quintero, and Schmidt}]{Cui:2020tdf}
\bibinfo{author}{Z.-F. Cui}, \bibinfo{author}{M.~Ding},
  \bibinfo{author}{F.~Gao}, \bibinfo{author}{K.~Raya},
  \bibinfo{author}{D.~Binosi}, \bibinfo{author}{L.~Chang},
  \bibinfo{author}{C.~D. Roberts},
  \bibinfo{author}{J.~Rodr\'{\i}guez-Quintero}, \bibinfo{author}{S.~M.
  Schmidt},
\newblock \bibinfo{title}{{Kaon and pion parton distributions}},
\newblock \bibinfo{journal}{Eur. Phys. J. C} \bibinfo{volume}{80}
  (\bibinfo{year}{2020}) \bibinfo{pages}{1064}.
\bibitem[{Yin et~al.(2023)Yin, Xu, Cui, Roberts, and
  Rodr\'\i{}guez-Quintero}]{Yin:2023dbw}
\bibinfo{author}{P.-L. Yin}, \bibinfo{author}{Y.-Z. Xu}, \bibinfo{author}{Z.-F.
  Cui}, \bibinfo{author}{C.~D. Roberts},
  \bibinfo{author}{J.~Rodr\'\i{}guez-Quintero},
\newblock \bibinfo{title}{{All-Orders Evolution of Parton Distributions:
  Principle, Practice, and Predictions}},
\newblock \bibinfo{journal}{Chin. Phys. Lett. \emph{Express}}
  \bibinfo{volume}{40} (\bibinfo{year}{2023}) \bibinfo{pages}{091201}.
\bibitem[{Brodsky et~al.(1998)Brodsky, Pauli, and Pinsky}]{Brodsky:1997de}
\bibinfo{author}{S.~J. Brodsky}, \bibinfo{author}{H.-C. Pauli},
  \bibinfo{author}{S.~S. Pinsky},
\newblock \bibinfo{title}{{Quantum chromodynamics and other field theories on
  the light cone}},
\newblock \bibinfo{journal}{Phys. Rept.} \bibinfo{volume}{301}
  (\bibinfo{year}{1998}) \bibinfo{pages}{299--486}.
\bibitem[{Heinzl(2001)}]{Heinzl:2000ht}
\bibinfo{author}{T.~Heinzl},
\newblock \bibinfo{title}{{Light cone quantization: Foundations and
  applications}},
\newblock \bibinfo{journal}{Lect. Notes Phys.} \bibinfo{volume}{572}
  (\bibinfo{year}{2001}) \bibinfo{pages}{55--142}.
\bibitem[{Xiao et~al.(2026)Xiao, Xu, Yao, Roberts, and
  Rodr{\'\i}guez-Quintero}]{Xiao:2025cqz}
\bibinfo{author}{Y.~Y. Xiao}, \bibinfo{author}{Z.~N. Xu},
  \bibinfo{author}{Z.~Q. Yao}, \bibinfo{author}{C.~D. Roberts},
  \bibinfo{author}{J.~Rodr{\'\i}guez-Quintero},
\newblock \bibinfo{title}{{Orbital angular momentum in the pion and kaon:
  Rest-frame and light-front}},
\newblock \bibinfo{journal}{Phys. Lett. B} \bibinfo{volume}{876}
  (\bibinfo{year}{2026}) \bibinfo{pages}{140361}.
\bibitem[{Gao et~al.(2014)Gao, Chang, Liu, Roberts, and Schmidt}]{Gao:2014bca}
\bibinfo{author}{F.~Gao}, \bibinfo{author}{L.~Chang}, \bibinfo{author}{Y.-X.
  Liu}, \bibinfo{author}{C.~D. Roberts}, \bibinfo{author}{S.~M. Schmidt},
\newblock \bibinfo{title}{{Parton distribution amplitudes of light vector
  mesons}},
\newblock \bibinfo{journal}{Phys. Rev. D} \bibinfo{volume}{90}
  (\bibinfo{year}{2014}) \bibinfo{pages}{014011}.
\bibitem[{Hilger et~al.(2017)Hilger, Gomez-Rocha, and
  Krassnigg}]{Hilger:2015ora}
\bibinfo{author}{T.~Hilger}, \bibinfo{author}{M.~Gomez-Rocha},
  \bibinfo{author}{A.~Krassnigg},
\newblock \bibinfo{title}{{Light-quarkonium spectra and
  orbital-angular-momentum decomposition in a
  Bethe\textendash{}Salpeter-equation approach}},
\newblock \bibinfo{journal}{Eur. Phys. J. C} \bibinfo{volume}{77}
  (\bibinfo{year}{2017}) \bibinfo{pages}{625}.
\bibitem[{Xu et~al.(2023)Xu, Yao, Qin, Cui, and Roberts}]{Xu:2022kng}
\bibinfo{author}{Z.-N. Xu}, \bibinfo{author}{Z.-Q. Yao}, \bibinfo{author}{S.-X.
  Qin}, \bibinfo{author}{Z.-F. Cui}, \bibinfo{author}{C.~D. Roberts},
\newblock \bibinfo{title}{{Bethe-Salpeter kernel and properties of
  strange-quark mesons}},
\newblock \bibinfo{journal}{Eur. Phys. J. A} \bibinfo{volume}{59}
  (\bibinfo{year}{2023}) \bibinfo{pages}{39}.
\bibitem[{Diehl(2003)}]{Diehl:2003ny}
\bibinfo{author}{M.~Diehl},
\newblock \bibinfo{title}{{Generalized parton distributions}},
\newblock \bibinfo{journal}{Phys. Rept.} \bibinfo{volume}{388}
  (\bibinfo{year}{2003}) \bibinfo{pages}{41--277}.
\bibitem[{Meissner et~al.(2008)Meissner, Metz, Schlegel, and
  Goeke}]{Meissner:2008ay}
\bibinfo{author}{S.~Meissner}, \bibinfo{author}{A.~Metz},
  \bibinfo{author}{M.~Schlegel}, \bibinfo{author}{K.~Goeke},
\newblock \bibinfo{title}{{Generalized parton correlation functions for a
  spin-0 hadron}},
\newblock \bibinfo{journal}{JHEP} \bibinfo{volume}{08} (\bibinfo{year}{2008})
  \bibinfo{pages}{038}.
\bibitem[{Zhang et~al.(2021)Zhang, Cui, Ping, and Roberts}]{Zhang:2020ecj}
\bibinfo{author}{J.-L. Zhang}, \bibinfo{author}{Z.-F. Cui},
  \bibinfo{author}{J.~Ping}, \bibinfo{author}{C.~D. Roberts},
\newblock \bibinfo{title}{{Contact interaction analysis of pion GTMDs}},
\newblock \bibinfo{journal}{Eur. Phys. J. C} \bibinfo{volume}{81}
  (\bibinfo{year}{2021}) \bibinfo{pages}{6}.
\bibitem[{Adhikari et~al.(2021)Adhikari, Mondal, Nair, Xu, Jia, Zhao, and
  Vary}]{Adhikari:2021jrh}
\bibinfo{author}{L.~Adhikari}, \bibinfo{author}{C.~Mondal},
  \bibinfo{author}{S.~Nair}, \bibinfo{author}{S.~Xu}, \bibinfo{author}{S.~Jia},
  \bibinfo{author}{X.~Zhao}, \bibinfo{author}{J.~P. Vary},
\newblock \bibinfo{title}{{Generalized parton distributions and spin structures
  of light mesons from a light-front Hamiltonian approach}},
\newblock \bibinfo{journal}{Phys. Rev. D} \bibinfo{volume}{104}
  (\bibinfo{year}{2021}) \bibinfo{pages}{114019}.
\bibitem[{Wang et~al.(2022)Wang, Xing, Kang, Raya, and Chang}]{Wang:2022mrh}
\bibinfo{author}{X.~Wang}, \bibinfo{author}{Z.~Xing},
  \bibinfo{author}{J.~Kang}, \bibinfo{author}{K.~Raya},
  \bibinfo{author}{L.~Chang},
\newblock \bibinfo{title}{{Pion scalar, vector, and tensor form factors from a
  contact interaction}},
\newblock \bibinfo{journal}{Phys. Rev. D} \bibinfo{volume}{106}
  (\bibinfo{year}{2022}) \bibinfo{pages}{054016}.
\bibitem[{Puhan and Dahiya(2025)}]{Puhan:2025pfs}
\bibinfo{author}{S.~Puhan}, \bibinfo{author}{H.~Dahiya},
\newblock \bibinfo{title}{{Scalar, vector, and tensor form factors of pion and
  kaon}},
\newblock \bibinfo{journal}{Phys. Rev. D} \bibinfo{volume}{111}
  (\bibinfo{year}{2025}) \bibinfo{pages}{114039}.
\bibitem[{Alexandrou et~al.(2022)Alexandrou, Bacchio, Cloet, Constantinou,
  Delmar, Hadjiyiannakou, Koutsou, Lauer, and Vaquero}]{Alexandrou:2021ztx}
\bibinfo{author}{C.~Alexandrou}, \bibinfo{author}{S.~Bacchio},
  \bibinfo{author}{I.~Cloet}, \bibinfo{author}{M.~Constantinou},
  \bibinfo{author}{J.~Delmar}, \bibinfo{author}{K.~Hadjiyiannakou},
  \bibinfo{author}{G.~Koutsou}, \bibinfo{author}{C.~Lauer},
  \bibinfo{author}{A.~Vaquero} (\bibinfo{collaboration}{ETM}),
\newblock \bibinfo{title}{{Scalar, vector, and tensor form factors for the pion
  and kaon from lattice QCD}},
\newblock \bibinfo{journal}{Phys. Rev. D} \bibinfo{volume}{105}
  (\bibinfo{year}{2022}) \bibinfo{pages}{054502}.
\bibitem[{Barabanov et~al.(2021)}]{Barabanov:2020jvn}
\bibinfo{author}{M.~Y. Barabanov}, et~al.,
\newblock \bibinfo{title}{{Diquark Correlations in Hadron Physics: Origin,
  Impact and Evidence}},
\newblock \bibinfo{journal}{Prog. Part. Nucl. Phys.} \bibinfo{volume}{116}
  (\bibinfo{year}{2021}) \bibinfo{pages}{103835}.
\bibitem[{Yin et~al.(2021)Yin, Cui, Roberts, and Segovia}]{Yin:2021uom}
\bibinfo{author}{P.-L. Yin}, \bibinfo{author}{Z.-F. Cui},
  \bibinfo{author}{C.~D. Roberts}, \bibinfo{author}{J.~Segovia},
\newblock \bibinfo{title}{{Masses of positive- and negative-parity hadron
  ground-states, including those with heavy quarks}},
\newblock \bibinfo{journal}{Eur. Phys. J. C} \bibinfo{volume}{81}
  (\bibinfo{year}{2021}) \bibinfo{pages}{327}.
\bibitem[{Eichmann et~al.(2016)Eichmann, Sanchis-Alepuz, Williams, Alkofer, and
  Fischer}]{Eichmann:2016yit}
\bibinfo{author}{G.~Eichmann}, \bibinfo{author}{H.~Sanchis-Alepuz},
  \bibinfo{author}{R.~Williams}, \bibinfo{author}{R.~Alkofer},
  \bibinfo{author}{C.~S. Fischer},
\newblock \bibinfo{title}{{Baryons as relativistic three-quark bound states}},
\newblock \bibinfo{journal}{Prog. Part. Nucl. Phys.} \bibinfo{volume}{91}
  (\bibinfo{year}{2016}) \bibinfo{pages}{1--100}.
\bibitem[{Fischer(2019)}]{Fischer:2018sdj}
\bibinfo{author}{C.~S. Fischer},
\newblock \bibinfo{title}{{QCD at finite temperature and chemical potential
  from Dyson--Schwinger equations}},
\newblock \bibinfo{journal}{Prog. Part. Nucl. Phys.} \bibinfo{volume}{105}
  (\bibinfo{year}{2019}) \bibinfo{pages}{1--60}.
\bibitem[{Huber(2020)}]{Huber:2018ned}
\bibinfo{author}{M.~Q. Huber},
\newblock \bibinfo{title}{{Nonperturbative properties of Yang-Mills theories}},
\newblock \bibinfo{journal}{Phys. Rept.} \bibinfo{volume}{879}
  (\bibinfo{year}{2020}) \bibinfo{pages}{1 -- 92}.
\bibitem[{Qin and Roberts(2020)}]{Qin:2020rad}
\bibinfo{author}{S.-X. Qin}, \bibinfo{author}{C.~D. Roberts},
\newblock \bibinfo{title}{{Impressions of the Continuum Bound State Problem in
  QCD}},
\newblock \bibinfo{journal}{Chin. Phys. Lett.} \bibinfo{volume}{37}
  (\bibinfo{year}{2020}) \bibinfo{pages}{121201}.
\bibitem[{Munczek(1995)}]{Munczek:1994zz}
\bibinfo{author}{H.~J. Munczek},
\newblock \bibinfo{title}{{Dynamical chiral symmetry breaking, Goldstone's
  theorem and the consistency of the Schwinger-Dyson and Bethe-Salpeter
  Equations}},
\newblock \bibinfo{journal}{Phys. Rev. D} \bibinfo{volume}{52}
  (\bibinfo{year}{1995}) \bibinfo{pages}{4736--4740}.
\bibitem[{Bender et~al.(1996)Bender, Roberts, and von Smekal}]{Bender:1996bb}
\bibinfo{author}{A.~Bender}, \bibinfo{author}{C.~D. Roberts},
  \bibinfo{author}{L.~von Smekal},
\newblock \bibinfo{title}{{Goldstone Theorem and Diquark Confinement Beyond
  Rainbow- Ladder Approximation}},
\newblock \bibinfo{journal}{Phys. Lett. B} \bibinfo{volume}{380}
  (\bibinfo{year}{1996}) \bibinfo{pages}{7--12}.
\bibitem[{Gutierrez-Guerrero et~al.(2010)Gutierrez-Guerrero, Bashir, Cloet, and
  Roberts}]{Gutierrez-Guerrero:2010waf}
\bibinfo{author}{L.~X. Gutierrez-Guerrero}, \bibinfo{author}{A.~Bashir},
  \bibinfo{author}{I.~C. Cloet}, \bibinfo{author}{C.~D. Roberts},
\newblock \bibinfo{title}{{Pion form factor from a contact interaction}},
\newblock \bibinfo{journal}{Phys. Rev. C} \bibinfo{volume}{81}
  (\bibinfo{year}{2010}) \bibinfo{pages}{065202}.
\bibitem[{Roberts et~al.(2011)Roberts, Bashir, Guti{\'e}rrez-Guerrero, Roberts,
  and Wilson}]{Roberts:2011wy}
\bibinfo{author}{H.~L.~L. Roberts}, \bibinfo{author}{A.~Bashir},
  \bibinfo{author}{L.~X. Guti{\'e}rrez-Guerrero}, \bibinfo{author}{C.~D.
  Roberts}, \bibinfo{author}{D.~J. Wilson},
\newblock \bibinfo{title}{{$\pi$- and $\rho$-mesons, and their diquark
  partners, from a contact interaction}},
\newblock \bibinfo{journal}{Phys. Rev. C} \bibinfo{volume}{83}
  (\bibinfo{year}{2011}) \bibinfo{pages}{065206}.
\bibitem[{Qin et~al.(2011)Qin, Chang, Liu, Roberts, and Wilson}]{Qin:2011dd}
\bibinfo{author}{S.-X. Qin}, \bibinfo{author}{L.~Chang}, \bibinfo{author}{Y.-X.
  Liu}, \bibinfo{author}{C.~D. Roberts}, \bibinfo{author}{D.~J. Wilson},
\newblock \bibinfo{title}{{Interaction model for the gap equation}},
\newblock \bibinfo{journal}{Phys. Rev. C} \bibinfo{volume}{84}
  (\bibinfo{year}{2011}) \bibinfo{pages}{042202(R)}.
\bibitem[{Binosi et~al.(2015)Binosi, Chang, Papavassiliou, and
  Roberts}]{Binosi:2014aea}
\bibinfo{author}{D.~Binosi}, \bibinfo{author}{L.~Chang},
  \bibinfo{author}{J.~Papavassiliou}, \bibinfo{author}{C.~D. Roberts},
\newblock \bibinfo{title}{{Bridging a gap between continuum-QCD and \emph{ab
  initio} predictions of hadron observables}},
\newblock \bibinfo{journal}{Phys. Lett. B} \bibinfo{volume}{742}
  (\bibinfo{year}{2015}) \bibinfo{pages}{183--188}.
\bibitem[{Chen et~al.(2013)Chen, Chang, Roberts, Wan, Schmidt, and
  Wilson}]{Chen:2012txa}
\bibinfo{author}{C.~Chen}, \bibinfo{author}{L.~Chang}, \bibinfo{author}{C.~D.
  Roberts}, \bibinfo{author}{S.-L. Wan}, \bibinfo{author}{S.~M. Schmidt},
  \bibinfo{author}{D.~J. Wilson},
\newblock \bibinfo{title}{{Features and flaws of a contact interaction
  treatment of the kaon}},
\newblock \bibinfo{journal}{Phys. Rev. C} \bibinfo{volume}{87}
  (\bibinfo{year}{2013}) \bibinfo{pages}{045207}.
\bibitem[{Serna et~al.(2017)Serna, El-Bennich, and Krein}]{Serna:2017nlr}
\bibinfo{author}{F.~E. Serna}, \bibinfo{author}{B.~El-Bennich},
  \bibinfo{author}{G.~Krein},
\newblock \bibinfo{title}{{Charmed mesons with a symmetry-preserving contact
  interaction}},
\newblock \bibinfo{journal}{Phys. Rev. D} \bibinfo{volume}{96}
  (\bibinfo{year}{2017}) \bibinfo{pages}{014013}.
\bibitem[{Xing et~al.(2022)Xing, Xu, Cui, Roberts, and Xu}]{Xing:2022sor}
\bibinfo{author}{H.-Y. Xing}, \bibinfo{author}{Z.-N. Xu},
  \bibinfo{author}{Z.-F. Cui}, \bibinfo{author}{C.~D. Roberts},
  \bibinfo{author}{C.~Xu},
\newblock \bibinfo{title}{{Heavy + heavy and heavy + light pseudoscalar to
  vector semileptonic transitions}},
\newblock \bibinfo{journal}{Eur. Phys. J. C} \bibinfo{volume}{82}
  (\bibinfo{year}{2022}) \bibinfo{pages}{889}.
\bibitem[{Sultan et~al.(2024)Sultan, Xing, Raya, Bashir, and
  Chang}]{Sultan:2024hep}
\bibinfo{author}{M.~A. Sultan}, \bibinfo{author}{Z.~Xing},
  \bibinfo{author}{K.~Raya}, \bibinfo{author}{A.~Bashir},
  \bibinfo{author}{L.~Chang},
\newblock \bibinfo{title}{{Gravitational form factors of pseudoscalar mesons in
  a contact interaction}},
\newblock \bibinfo{journal}{Phys. Rev. D} \bibinfo{volume}{110}
  (\bibinfo{year}{2024}) \bibinfo{pages}{054034}.
\bibitem[{Xing et~al.(2025)Xing, Bian, Cui, and Roberts}]{Xing:2025eip}
\bibinfo{author}{H.-Y. Xing}, \bibinfo{author}{W.-H. Bian},
  \bibinfo{author}{Z.-F. Cui}, \bibinfo{author}{C.~D. Roberts},
\newblock \bibinfo{title}{{Kaon and pion fragmentation functions}},
\newblock \bibinfo{journal}{Eur. Phys. J. C} \bibinfo{volume}{85}
  (\bibinfo{year}{2025}) \bibinfo{pages}{1305}.
\bibitem[{Guti{\'e}rrez-Guerrero et~al.(2019)Guti{\'e}rrez-Guerrero, Bashir,
  Bedolla, and Santopinto}]{Gutierrez-Guerrero:2019uwa}
\bibinfo{author}{L.~Guti{\'e}rrez-Guerrero}, \bibinfo{author}{A.~Bashir},
  \bibinfo{author}{M.~A. Bedolla}, \bibinfo{author}{E.~Santopinto},
\newblock \bibinfo{title}{{Masses of Light and Heavy Mesons and Baryons: A
  Unified Picture}},
\newblock \bibinfo{journal}{Phys. Rev. D} \bibinfo{volume}{100}
  (\bibinfo{year}{2019}) \bibinfo{pages}{114032}.
\bibitem[{Chen et~al.(2025)Chen, Gao, and Qin}]{Chen:2024emt}
\bibinfo{author}{C.~Chen}, \bibinfo{author}{F.~Gao}, \bibinfo{author}{S.-X.
  Qin},
\newblock \bibinfo{title}{{Screening masses of positive- and negative-parity
  hadron ground states, including those with strangeness}},
\newblock \bibinfo{journal}{Phys. Rev. D} \bibinfo{volume}{112}
  (\bibinfo{year}{2025}) \bibinfo{pages}{014022}.
\bibitem[{Cheng et~al.(2026)Cheng, Ding, Binosi, and Roberts}]{Cheng:2026nud}
\bibinfo{author}{D.-D. Cheng}, \bibinfo{author}{M.~Ding},
  \bibinfo{author}{D.~Binosi}, \bibinfo{author}{C.~D. Roberts},
\newblock \bibinfo{title}{{Kaon Boer-Mulders function using a contact
  interaction -- arXiv:2603.25941 [hep-ph]}},
\newblock \bibinfo{journal}{Eur. Phys. J. C}  (\bibinfo{year}{2026})
  \bibinfo{pages}{\emph{in press}}.
\bibitem[{Guti\'{e}rrez-Guerrero and
  Hern\'{a}ndez-Pinto(2026)}]{Gutierrez-Guerrero:2026rsb}
\bibinfo{author}{L.~X. Guti\'{e}rrez-Guerrero}, \bibinfo{author}{R.~J.
  Hern\'{a}ndez-Pinto},
\newblock \bibinfo{title}{Symmetry-preserving contact interaction approaches:
  An overview of meson and diquark form factors},
\newblock \bibinfo{journal}{Particles} \bibinfo{volume}{9}
  (\bibinfo{year}{2026}) \bibinfo{pages}{45}.
\bibitem[{Segovia et~al.(2013)Segovia, Chen, Roberts, and
  Wan}]{Segovia:2013rca}
\bibinfo{author}{J.~Segovia}, \bibinfo{author}{C.~Chen}, \bibinfo{author}{C.~D.
  Roberts}, \bibinfo{author}{S.-L. Wan},
\newblock \bibinfo{title}{{Insights into the {$\gamma^\ast N \to \Delta$}
  transition}},
\newblock \bibinfo{journal}{Phys. Rev. C} \bibinfo{volume}{88}
  (\bibinfo{year}{2013}) \bibinfo{pages}{032201(R)}.
\bibitem[{Xu et~al.(2015)Xu, Chen, Cloet, Roberts, Segovia, and
  Zong}]{Xu:2015kta}
\bibinfo{author}{S.-S. Xu}, \bibinfo{author}{C.~Chen}, \bibinfo{author}{I.~C.
  Cloet}, \bibinfo{author}{C.~D. Roberts}, \bibinfo{author}{J.~Segovia},
  \bibinfo{author}{H.-S. Zong},
\newblock \bibinfo{title}{{Contact-interaction Faddeev equation and,
  \emph{inter alia}, proton tensor charges}},
\newblock \bibinfo{journal}{Phys. Rev. D} \bibinfo{volume}{92}
  (\bibinfo{year}{2015}) \bibinfo{pages}{114034}.
\bibitem[{Raya et~al.(2021)Raya, Guti\'errez-Guerrero, Bashir, Chang, Cui, Lu,
  Roberts, and Segovia}]{Raya:2021pyr}
\bibinfo{author}{K.~Raya}, \bibinfo{author}{L.~X. Guti\'errez-Guerrero},
  \bibinfo{author}{A.~Bashir}, \bibinfo{author}{L.~Chang},
  \bibinfo{author}{Z.~F. Cui}, \bibinfo{author}{Y.~Lu}, \bibinfo{author}{C.~D.
  Roberts}, \bibinfo{author}{J.~Segovia},
\newblock \bibinfo{title}{{Dynamical diquarks in the \mbox{$\gamma^{(\ast)}
  p\to N(1535)\tfrac{1}{2}^-$} transition}},
\newblock \bibinfo{journal}{Eur. Phys. J. A} \bibinfo{volume}{57}
  (\bibinfo{year}{2021}) \bibinfo{pages}{266}.
\bibitem[{Cheng et~al.(2022)Cheng, Serna, Yao, Chen, Cui, and
  Roberts}]{Cheng:2022jxe}
\bibinfo{author}{P.~Cheng}, \bibinfo{author}{F.~E. Serna},
  \bibinfo{author}{Z.-Q. Yao}, \bibinfo{author}{C.~Chen},
  \bibinfo{author}{Z.-F. Cui}, \bibinfo{author}{C.~D. Roberts},
\newblock \bibinfo{title}{{Contact interaction analysis of octet baryon
  axial-vector and pseudoscalar form factors}},
\newblock \bibinfo{journal}{Phys. Rev. D} \bibinfo{volume}{106}
  (\bibinfo{year}{2022}) \bibinfo{pages}{054031}.
\bibitem[{Yu et~al.(2025)Yu, Cheng, Xing, Binosi, and Roberts}]{Yu:2025fer}
\bibinfo{author}{Y.~Yu}, \bibinfo{author}{P.~Cheng}, \bibinfo{author}{H.-Y.
  Xing}, \bibinfo{author}{D.~Binosi}, \bibinfo{author}{C.~D. Roberts},
\newblock \bibinfo{title}{{Distribution Functions of $\Lambda$ and $\Sigma^0$
  Baryons}},
\newblock \bibinfo{journal}{Eur. Phys. J. A} \bibinfo{volume}{61}
  (\bibinfo{year}{2025}) \bibinfo{pages}{208}.
\bibitem[{Bai et~al.(2026)Bai, Lu, Yao, Roberts, and Schmidt}]{Bai:2026nqo}
\bibinfo{author}{X.-Y. Bai}, \bibinfo{author}{Y.~Lu}, \bibinfo{author}{Z.-Q.
  Yao}, \bibinfo{author}{C.~D. Roberts}, \bibinfo{author}{S.~M. Schmidt},
\newblock \bibinfo{title}{{Contact interaction treatment of the nucleon Faddeev
  equation -- arXiv:2602.02880 [hep-ph]}}  (\bibinfo{year}{2026}).
\bibitem[{Jaffe and Ross(1980)}]{Jaffe:1980ti}
\bibinfo{author}{R.~L. Jaffe}, \bibinfo{author}{G.~G. Ross},
\newblock \bibinfo{title}{{Normalizing the Renormalization Group Analysis of
  Deep Inelastic Leptoproduction}},
\newblock \bibinfo{journal}{Phys. Lett. B} \bibinfo{volume}{93}
  (\bibinfo{year}{1980}) \bibinfo{pages}{313--317}.
\bibitem[{Dokshitzer(1977)}]{Dokshitzer:1977sg}
\bibinfo{author}{Y.~L. Dokshitzer},
\newblock \bibinfo{title}{Calculation of the structure functions for deep
  inelastic scattering and $e^+$ $e^-$ annihilation by perturbation theory in
  quantum chromodynamics. ({\mbox {i}n {r}ussian})},
\newblock \bibinfo{journal}{Sov. Phys. JETP} \bibinfo{volume}{46}
  (\bibinfo{year}{1977}) \bibinfo{pages}{641--653}.
\bibitem[{Gribov and Lipatov(1971)}]{Gribov:1971zn}
\bibinfo{author}{V.~N. Gribov}, \bibinfo{author}{L.~N. Lipatov},
\newblock \bibinfo{title}{{Deep inelastic electron scattering in perturbation
  theory}},
\newblock \bibinfo{journal}{Phys. Lett. B} \bibinfo{volume}{37}
  (\bibinfo{year}{1971}) \bibinfo{pages}{78--80}.
\bibitem[{Lipatov(1975)}]{Lipatov:1974qm}
\bibinfo{author}{L.~N. Lipatov},
\newblock \bibinfo{title}{{The parton model and perturbation theory}},
\newblock \bibinfo{journal}{Sov. J. Nucl. Phys.} \bibinfo{volume}{20}
  (\bibinfo{year}{1975}) \bibinfo{pages}{94--102}.
\bibitem[{Altarelli and Parisi(1977)}]{Altarelli:1977zs}
\bibinfo{author}{G.~Altarelli}, \bibinfo{author}{G.~Parisi},
\newblock \bibinfo{title}{{Asymptotic Freedom in Parton Language}},
\newblock \bibinfo{journal}{Nucl. Phys. B} \bibinfo{volume}{126}
  (\bibinfo{year}{1977}) \bibinfo{pages}{298--318}.
\bibitem[{Yamanaka et~al.(2013)Yamanaka, Doi, Imai, and
  Suganuma}]{Yamanaka:2013zoa}
\bibinfo{author}{N.~Yamanaka}, \bibinfo{author}{T.~M. Doi},
  \bibinfo{author}{S.~Imai}, \bibinfo{author}{H.~Suganuma},
\newblock \bibinfo{title}{{Quark tensor charge and electric dipole moment
  within the Schwinger-Dyson formalism}},
\newblock \bibinfo{journal}{Phys. Rev. D} \bibinfo{volume}{88}
  (\bibinfo{year}{2013}) \bibinfo{pages}{074036}.
\bibitem[{Wang et~al.(2018)Wang, Qin, Roberts, and Schmidt}]{Wang:2018kto}
\bibinfo{author}{Q.-W. Wang}, \bibinfo{author}{S.-X. Qin},
  \bibinfo{author}{C.~D. Roberts}, \bibinfo{author}{S.~M. Schmidt},
\newblock \bibinfo{title}{{Proton tensor charges from a Poincar{\'e}-covariant
  Faddeev equation}},
\newblock \bibinfo{journal}{Phys. Rev. D} \bibinfo{volume}{98}
  (\bibinfo{year}{2018}) \bibinfo{pages}{054019}.
\bibitem[{Xing and Chang(2023)}]{Xing:2022jtt}
\bibinfo{author}{Z.~Xing}, \bibinfo{author}{L.~Chang},
\newblock \bibinfo{title}{{Symmetry preserving contact interaction treatment of
  the kaon}},
\newblock \bibinfo{journal}{Phys. Rev. D} \bibinfo{volume}{107}
  (\bibinfo{year}{2023}) \bibinfo{pages}{014019}.
\bibitem[{Pitschmann et~al.(2015)Pitschmann, Seng, Roberts, and
  Schmidt}]{Pitschmann:2014jxa}
\bibinfo{author}{M.~Pitschmann}, \bibinfo{author}{C.-Y. Seng},
  \bibinfo{author}{C.~D. Roberts}, \bibinfo{author}{S.~M. Schmidt},
\newblock \bibinfo{title}{{Nucleon tensor charges and electric dipole
  moments}},
\newblock \bibinfo{journal}{Phys. Rev. D} \bibinfo{volume}{91}
  (\bibinfo{year}{2015}) \bibinfo{pages}{074004}.
\bibitem[{Cui et~al.(2020)Cui, Zhang, Binosi, de~Soto, Mezrag, Papavassiliou,
  Roberts, Rodr{\'{\i}}guez-Quintero, Segovia, and Zafeiropoulos}]{Cui:2019dwv}
\bibinfo{author}{Z.-F. Cui}, \bibinfo{author}{J.-L. Zhang},
  \bibinfo{author}{D.~Binosi}, \bibinfo{author}{F.~de~Soto},
  \bibinfo{author}{C.~Mezrag}, \bibinfo{author}{J.~Papavassiliou},
  \bibinfo{author}{C.~D. Roberts},
  \bibinfo{author}{J.~Rodr{\'{\i}}guez-Quintero}, \bibinfo{author}{J.~Segovia},
  \bibinfo{author}{S.~Zafeiropoulos},
\newblock \bibinfo{title}{{Effective charge from lattice QCD}},
\newblock \bibinfo{journal}{Chin. Phys. C} \bibinfo{volume}{44}
  (\bibinfo{year}{2020}) \bibinfo{pages}{083102}.
\bibitem[{Deur et~al.(2024)Deur, Brodsky, and Roberts}]{Deur:2023dzc}
\bibinfo{author}{A.~Deur}, \bibinfo{author}{S.~J. Brodsky},
  \bibinfo{author}{C.~D. Roberts},
\newblock \bibinfo{title}{{QCD Running Couplings and Effective Charges}},
\newblock \bibinfo{journal}{Prog. Part. Nucl. Phys.} \bibinfo{volume}{134}
  (\bibinfo{year}{2024}) \bibinfo{pages}{104081}.
\bibitem[{Brodsky et~al.(2024)Brodsky, Deur, and Roberts}]{Brodsky:2024zev}
\bibinfo{author}{S.~J. Brodsky}, \bibinfo{author}{A.~Deur},
  \bibinfo{author}{C.~D. Roberts},
\newblock \bibinfo{title}{{The Secret to the Strongest Force in the Universe}},
\newblock \bibinfo{journal}{Sci. Am.} \bibinfo{volume}{5 (May)}
  (\bibinfo{year}{2024}) \bibinfo{pages}{32--39}.
\bibitem[{He and Ji(1995)}]{He:1994gz}
\bibinfo{author}{H.~He}, \bibinfo{author}{X.~Ji},
\newblock \bibinfo{title}{{The Nucleon's tensor charge}},
\newblock \bibinfo{journal}{Phys. Rev. D} \bibinfo{volume}{52}
  (\bibinfo{year}{1995}) \bibinfo{pages}{2960--2963}.
\bibitem[{Bhattacharya et~al.(2015)Bhattacharya, Cirigliano, Cohen, Gupta,
  Joseph, Lin, and Yoon}]{Bhattacharya:2015wna}
\bibinfo{author}{T.~Bhattacharya}, \bibinfo{author}{V.~Cirigliano},
  \bibinfo{author}{S.~Cohen}, \bibinfo{author}{R.~Gupta},
  \bibinfo{author}{A.~Joseph}, \bibinfo{author}{H.-W. Lin},
  \bibinfo{author}{B.~Yoon} (\bibinfo{collaboration}{PNDME}),
\newblock \bibinfo{title}{{Iso-vector and Iso-scalar Tensor Charges of the
  Nucleon from Lattice QCD}},
\newblock \bibinfo{journal}{Phys. Rev. D} \bibinfo{volume}{92}
  (\bibinfo{year}{2015}) \bibinfo{pages}{094511}.
\bibitem[{Abdel-Rehim et~al.(2015)}]{Abdel-Rehim:2015owa}
\bibinfo{author}{A.~Abdel-Rehim}, et~al.,
\newblock \bibinfo{title}{{Nucleon and pion structure with lattice QCD
  simulations at physical value of the pion mass}},
\newblock \bibinfo{journal}{Phys. Rev. D} \bibinfo{volume}{92}
  (\bibinfo{year}{2015}) \bibinfo{pages}{114513}. \bibinfo{note}{[Erratum:
  Phys. Rev. D 93, 039904 (2016)]}.
\bibitem[{Alexandrou(2024)}]{Alexandrou:2024awx}
\bibinfo{author}{C.~Alexandrou},
\newblock \bibinfo{title}{{Nucleon Transversity from lattice QCD}},
\newblock \bibinfo{journal}{PoS} \bibinfo{volume}{Transversity2024}
  (\bibinfo{year}{2024}) \bibinfo{pages}{002}.
\bibitem[{Ye et~al.(2017)Ye, Sato, Allada, Liu, Chen, Gao, Kang, Prokudin, Sun,
  and Yuan}]{Ye:2016prn}
\bibinfo{author}{Z.~Ye}, \bibinfo{author}{N.~Sato},
  \bibinfo{author}{K.~Allada}, \bibinfo{author}{T.~Liu}, \bibinfo{author}{J.-P.
  Chen}, \bibinfo{author}{H.~Gao}, \bibinfo{author}{Z.-B. Kang},
  \bibinfo{author}{A.~Prokudin}, \bibinfo{author}{P.~Sun},
  \bibinfo{author}{F.~Yuan},
\newblock \bibinfo{title}{{Unveiling the nucleon tensor charge at Jefferson
  Lab: A study of the SoLID case}},
\newblock \bibinfo{journal}{Phys. Lett. B} \bibinfo{volume}{767}
  (\bibinfo{year}{2017}) \bibinfo{pages}{91--98}.
\bibitem[{Cui et~al.(2022)Cui, Binosi, Roberts, and Schmidt}]{Cui:2022fyr}
\bibinfo{author}{Z.-F. Cui}, \bibinfo{author}{D.~Binosi},
  \bibinfo{author}{C.~D. Roberts}, \bibinfo{author}{S.~M. Schmidt},
\newblock \bibinfo{title}{{Hadron and light nucleus radii from electron
  scattering}},
\newblock \bibinfo{journal}{Chin. Phys. C} \bibinfo{volume}{46}
  (\bibinfo{year}{2022}) \bibinfo{pages}{122001}.
\bibitem[{Hoferichter et~al.(2019)Hoferichter, Kubis, Ruiz~de Elvira, and
  Stoffer}]{Hoferichter:2018zwu}
\bibinfo{author}{M.~Hoferichter}, \bibinfo{author}{B.~Kubis},
  \bibinfo{author}{J.~Ruiz~de Elvira}, \bibinfo{author}{P.~Stoffer},
\newblock \bibinfo{title}{{Nucleon Matrix Elements of the Antisymmetric Quark
  Tensor}},
\newblock \bibinfo{journal}{Phys. Rev. Lett.} \bibinfo{volume}{122}
  (\bibinfo{year}{2019}) \bibinfo{pages}{122001}. \bibinfo{note}{[Erratum:
  Phys. Rev. Lett. 124, 199901 (2020)]}.
\bibitem[{Cosyn and Pire(2018)}]{Cosyn:2018rdm}
\bibinfo{author}{W.~Cosyn}, \bibinfo{author}{B.~Pire},
\newblock \bibinfo{title}{{Transversity generalized parton distributions for
  the deuteron}},
\newblock \bibinfo{journal}{Phys. Rev. D} \bibinfo{volume}{98}
  (\bibinfo{year}{2018}) \bibinfo{pages}{074020}.
\bibitem[{Jarecke et~al.(2003)Jarecke, Maris, and Tandy}]{Jarecke:2002xd}
\bibinfo{author}{D.~Jarecke}, \bibinfo{author}{P.~Maris},
  \bibinfo{author}{P.~C. Tandy},
\newblock \bibinfo{title}{{Strong decays of light vector mesons}},
\newblock \bibinfo{journal}{Phys. Rev. C} \bibinfo{volume}{67}
  (\bibinfo{year}{2003}) \bibinfo{pages}{035202}.
\bibitem[{Williams et~al.(2016)Williams, Fischer, and
  Heupel}]{Williams:2015cvx}
\bibinfo{author}{R.~Williams}, \bibinfo{author}{C.~S. Fischer},
  \bibinfo{author}{W.~Heupel},
\newblock \bibinfo{title}{{Light mesons in QCD and unquenching effects from the
  3PI effective action}},
\newblock \bibinfo{journal}{Phys. Rev. D} \bibinfo{volume}{93}
  (\bibinfo{year}{2016}) \bibinfo{pages}{034026}.
\bibitem[{Xu et~al.(2021)Xu, Chen, Yao, Binosi, Cui, and Roberts}]{Xu:2021mju}
\bibinfo{author}{Y.-Z. Xu}, \bibinfo{author}{S.~Chen}, \bibinfo{author}{Z.-Q.
  Yao}, \bibinfo{author}{D.~Binosi}, \bibinfo{author}{Z.-F. Cui},
  \bibinfo{author}{C.~D. Roberts},
\newblock \bibinfo{title}{{Vector-meson production and vector meson
  dominance}},
\newblock \bibinfo{journal}{Eur. Phys. J. C} \bibinfo{volume}{81}
  (\bibinfo{year}{2021}) \bibinfo{pages}{895}.
\bibitem[{Shi et~al.(2022)Shi, Li, Li, Chen, and Jia}]{Shi:2022erw}
\bibinfo{author}{C.~Shi}, \bibinfo{author}{J.~Li}, \bibinfo{author}{M.~Li},
  \bibinfo{author}{X.~Chen}, \bibinfo{author}{W.~Jia},
\newblock \bibinfo{title}{{Transverse momentum distributions of valence quarks
  in light and heavy vector mesons}},
\newblock \bibinfo{journal}{Phys. Rev. D} \bibinfo{volume}{106}
  (\bibinfo{year}{2022}) \bibinfo{pages}{014026}.
\bibitem[{Kaur et~al.(2024)Kaur, Wu, Hu, Lan, Mondal, Zhao, and
  Vary}]{Kaur:2024iwn}
\bibinfo{author}{S.~Kaur}, \bibinfo{author}{J.~Wu}, \bibinfo{author}{Z.~Hu},
  \bibinfo{author}{J.~Lan}, \bibinfo{author}{C.~Mondal},
  \bibinfo{author}{X.~Zhao}, \bibinfo{author}{J.~P. Vary}
  (\bibinfo{collaboration}{BLFQ}),
\newblock \bibinfo{title}{{Quark and gluon distributions in
  {\ensuremath{\rho}}-meson from basis light-front quantization}},
\newblock \bibinfo{journal}{Phys. Lett. B} \bibinfo{volume}{851}
  (\bibinfo{year}{2024}) \bibinfo{pages}{138563}.
\bibitem[{Zhang and Wu(2025)}]{Zhang:2024plq}
\bibinfo{author}{J.-L. Zhang}, \bibinfo{author}{J.~Wu},
\newblock \bibinfo{title}{{$\rho $ meson transverse momentum-dependent parton
  distributions}},
\newblock \bibinfo{journal}{Eur. Phys. J. C} \bibinfo{volume}{85}
  (\bibinfo{year}{2025}) \bibinfo{pages}{13}.
\bibitem[{Liu and Zahed(2025)}]{Liu:2025fuf}
\bibinfo{author}{W.-Y. Liu}, \bibinfo{author}{I.~Zahed},
\newblock \bibinfo{title}{{Tomography of the rho meson in the QCD instanton
  vacuum: Transverse momentum dependent parton distribution functions}},
\newblock \bibinfo{journal}{Phys. Rev. D} \bibinfo{volume}{112}
  (\bibinfo{year}{2025}) \bibinfo{pages}{034028}.
\bibitem[{Cahill et~al.(1987)Cahill, Roberts, and Praschifka}]{Cahill:1987qr}
\bibinfo{author}{R.~T. Cahill}, \bibinfo{author}{C.~D. Roberts},
  \bibinfo{author}{J.~Praschifka},
\newblock \bibinfo{title}{{Calculation of diquark masses in QCD}},
\newblock \bibinfo{journal}{Phys. Rev. D} \bibinfo{volume}{36}
  (\bibinfo{year}{1987}) \bibinfo{pages}{2804}.
\bibitem[{Roberts et~al.(2011)Roberts, Chang, Cloet, and
  Roberts}]{Roberts:2011cf}
\bibinfo{author}{H.~L.~L. Roberts}, \bibinfo{author}{L.~Chang},
  \bibinfo{author}{I.~C. Cloet}, \bibinfo{author}{C.~D. Roberts},
\newblock \bibinfo{title}{{Masses of ground and excited-state hadrons}},
\newblock \bibinfo{journal}{Few Body Syst.} \bibinfo{volume}{51}
  (\bibinfo{year}{2011}) \bibinfo{pages}{1--25}.
\bibitem[{Gao et~al.(2018)Gao, Qin, Roberts, and
  Rodr{\'{\i}}guez-Quintero}]{Gao:2017uox}
\bibinfo{author}{F.~Gao}, \bibinfo{author}{S.-X. Qin}, \bibinfo{author}{C.~D.
  Roberts}, \bibinfo{author}{J.~Rodr{\'{\i}}guez-Quintero},
\newblock \bibinfo{title}{{Locating the Gribov horizon}},
\newblock \bibinfo{journal}{Phys. Rev. D} \bibinfo{volume}{97}
  (\bibinfo{year}{2018}) \bibinfo{pages}{034010}.
\bibitem[{Ebert et~al.(1996)Ebert, Feldmann, and Reinhardt}]{Ebert:1996vx}
\bibinfo{author}{D.~Ebert}, \bibinfo{author}{T.~Feldmann},
  \bibinfo{author}{H.~Reinhardt},
\newblock \bibinfo{title}{{Extended NJL model for light and heavy mesons
  without $q \bar q$ thresholds}},
\newblock \bibinfo{journal}{Phys. Lett. B} \bibinfo{volume}{388}
  (\bibinfo{year}{1996}) \bibinfo{pages}{154--160}.
\bibitem[{Roberts et~al.(1992)Roberts, Williams, and Krein}]{Krein:1990sf}
\bibinfo{author}{C.~D. Roberts}, \bibinfo{author}{A.~G. Williams},
  \bibinfo{author}{G.~Krein},
\newblock \bibinfo{title}{{On the implications of confinement}},
\newblock \bibinfo{journal}{Int. J. Mod. Phys. A} \bibinfo{volume}{7}
  (\bibinfo{year}{1992}) \bibinfo{pages}{5607--5624}.
\bibitem[{Llewellyn-Smith(1969)}]{LlewellynSmith:1969az}
\bibinfo{author}{C.~H. Llewellyn-Smith},
\newblock \bibinfo{title}{{A relativistic formulation for the quark model for
  mesons}},
\newblock \bibinfo{journal}{Annals Phys.} \bibinfo{volume}{53}
  (\bibinfo{year}{1969}) \bibinfo{pages}{521--558}.
\bibitem[{Krassnigg and Blank(2011)}]{Krassnigg:2010mh}
\bibinfo{author}{A.~Krassnigg}, \bibinfo{author}{M.~Blank},
\newblock \bibinfo{title}{{A covariant study of tensor mesons}},
\newblock \bibinfo{journal}{Phys. Rev. D} \bibinfo{volume}{83}
  (\bibinfo{year}{2011}) \bibinfo{pages}{096006}.
\bibitem[{Maris and Tandy(1999)}]{Maris:1999nt}
\bibinfo{author}{P.~Maris}, \bibinfo{author}{P.~C. Tandy},
\newblock \bibinfo{title}{{Bethe-Salpeter study of vector meson masses and
  decay constants}},
\newblock \bibinfo{journal}{Phys. Rev. C} \bibinfo{volume}{60}
  (\bibinfo{year}{1999}) \bibinfo{pages}{055214}.
\bibitem[{Maris and Roberts(1997)}]{Maris:1997tm}
\bibinfo{author}{P.~Maris}, \bibinfo{author}{C.~D. Roberts},
\newblock \bibinfo{title}{{{$\pi$} and {$K$} meson Bethe-Salpeter amplitudes}},
\newblock \bibinfo{journal}{Phys. Rev. C} \bibinfo{volume}{56}
  (\bibinfo{year}{1997}) \bibinfo{pages}{3369--3383}.

\end{thebibliography}

\end{document}